\RequirePackage[2020-02-02]{latexrelease}
\documentclass[aps,prx,twocolumn,reprint]{revtex4-2}

\usepackage[utf8]{inputenc}
\usepackage{graphicx}
\usepackage{comment,xcolor}
\usepackage{amsmath,mathtools,amssymb,bm}
\usepackage{braket}
\usepackage{hyperref}

\newcommand{\mL}{\mathcal{L}}

\newcommand{\mR}{\mathcal{R}}
\newcommand{\mV}{\mathcal{V}}

\newcommand{\mA}{\mathcal{A}}

\newcommand{\Tr}{\text{Tr}}

\newcommand{\Htot}{H_\text{tot}}

\newcommand{\mtc}{m_\text{tc}}

\newcommand{\tgamma}{\tilde{\gamma}}

\newcommand{\mU}{\mathcal{U}}
\newcommand{\mP}{\mathcal{P}}
\newcommand{\mT}{\mathcal{T}}
\newcommand{\MF}{\text{MF}}

\newcommand{\ua}{\uparrow}
\newcommand{\da}{\downarrow}

\begin{document}

\title{Criticality and Rigidity of Dissipative Discrete Time Crystals in Solids}

\author{Koki Chinzei}
\affiliation{Institute for Solid State Physics, University of Tokyo, Kashiwa, Chiba 277-8581, Japan}

\author{Tatsuhiko N. Ikeda}
\affiliation{Institute for Solid State Physics, University of Tokyo, Kashiwa, Chiba 277-8581, Japan}

\date{\today}

\begin{abstract}
    We consider a dissipative quantum Ising model periodically driven by a train of $\pi$-pulses and investigate dissipative discrete time crystals (DTCs) in solids.
    In this model, the interaction between the spins spontaneously breaks the discrete time translation symmetry, giving rise to a dissipative DTC,
    where two ferromagnetic states are switched alternately by each pulse.
    We microscopically describe the generic dissipation due to thermal contact to an equilibrium heat bath using the Bloch-Redfield equation.
    In contrast to other DTC studies, this dissipation stabilizes, rather than destroys, the DTC order without fine-tuning as long as the temperature is low enough.
    Invoking the time-dependent mean-field theory and solving self-consistently the periodic drive, dissipation, and DTC order parameter, we investigate the nonequilibrium DTC phase transition and determine the critical exponents, including a dynamical one.
    We also find phase transitions without equilibrium counterpart: a nontrivial interplay of the periodic drive and dissipation gives rise to reentrant DTC transition when changing the pulse interval at a fixed temperature.
    Besides, to demonstrate the rigidity of the DTC, we consider imperfect $\pi$-pulses, showing that the DTC is robust against the small imperfections and finding that discrete time quasicrystals (DTQC) can appear for the larger imperfections. 
    Together with experimental proposals in magnetic materials, our results pave the ways for realizing the DTC and for uncovering nonequilibrium critical phenomena in real solid-state materials.
\end{abstract}

\maketitle


\section{Introduction}

Phase transition and criticality are key concepts for understanding many-body quantum physics,
and a general theoretical framework has been formulated to describe and classify the universal aspects of critical phenomena for equilibrium states of matter~\cite{Cardy1996}.
Recently, several theoretical studies have found exotic phases of matter without equilibrium counterparts in nonequilibrium conditions, such as many-body localization~\cite{Abanin2019} and Floquet topological phases~\cite{Oka2009,Kitagawa2010,Kitagawa2011,Jiang2011,Rudner2013,Potter2016,Kolodrubetz2018}.
The state-of-the-art technologies enable addressing such nonequilibrium phases experimentally~\cite{Schreiber2015,Eckardt2017,McIver2020}.

A discrete time crystal (DTC) is a genuinely nonequilibrium phase occurring in periodically-driven quantum (Floquet) systems~\cite{Else2016,VonKeyserlingk2016,Yao2017,Else2017,Zeng2017,Machado2020,Luitz2020, Sacha2015,Sacha2017b,Russomanno2017, Ho2017, Mizuta2018, Yu2019b, Giergiel2019b, Zhao2019,Pizzi2019, Collado2021} as the time-crystalline behavior is prohibited in equilibrium conditions~\cite{Wilczek2012,Li2012,Bruno2013,Bruno2013a,Watanabe2015,Khemani2019}
(Note, however, that some theoretical studies suggest that equilibrium time crystals are possible if one drops the requirements of rigidity and spatial infinite range order~\cite{Buca2019, Buca2019c, Medenjak2019, Dogra2019}, or locality~\cite{Kozin2019}).
The DTC is characterized by a breakdown of the discrete time translation symmetry entailing
subharmonic oscillations with period $n\tau$, where $\tau$ is the Floquet period and $n=2,3,\cdots$.

In idealistic dissipationless cases, the DTC was theoretically proposed in many systems such as many-body localized systems~\cite{Else2016,VonKeyserlingk2016,Yao2017}, prethermal systems~\cite{Else2017,Zeng2017,Machado2020,Luitz2020}, and so on~\cite{Sacha2015,Sacha2017b,Russomanno2017, Ho2017, Mizuta2018, Yu2019b, Giergiel2019b, Zhao2019,Pizzi2019, Collado2021}.
In theses systems, the Floquet heating~\cite{DAlessio2014,Lazarides2014,Kim2014}, which generically prevents the realization of the DTC in many-body quantum systems, is suppressed in some ways.
Furthermore, the dissipationless DTC has been experimentally demonstrated in cold-atom systems~\cite{Bordia2017, Zhang2017, Pal2018, Rovny2018a, Rovny2018}, quantum computers~\cite{Ippoliti2020, Randall2021, Mi2021, Frey2021}, and NV centers in diamond~\cite{Choi2017}.
In experiments, small dissipation and decoherence are always unavoidable and usually destroy the DTC behavior~\cite{Lazarides2017}.
Thus, these experiments are interpreted to have witnessed the DTC as a transient state, which would vanish in the long run.

Meanwhile, a new type of time crystal, the dissipative time crystal, has been proposed in some special models, with both continuous~\cite{Buca2019} and discrete~\cite{Gong2018} time evolution. 
The dissipative time crystal is a many-body system in which the time crystalline order is stabilized by dissipation rather than destroyed, surviving even in the long-time limit. 
Specifically, in the Floquet system, dissipation cools the system and suppresses the heating, stabilizing the DTC order.
Such a dissipative DTC has attracted lots of attention~\cite{Lledo2019, Lazarides2020, Riera2020, Chinzei2020, Gambetta2019a,Gambetta2019b, OSullivan2020} and has been experimentally verified in a cavity QED system recently~\cite{Kebler2021}.

However, so far, theories and experiments of the quantum dissipative DTCs have been limited to well-designed artificial quantum systems,
where one could control dissipation as well as the Hamiltonian.
On the other hand, if the DTC is realized in generic solid-state materials, it could offer a new way of controlling nonequilibrium phases of matter in material science.
However, solid-state systems are coupled to many external degrees of freedom like phonons, which cannot be fine-tuned and leads to more complicated dissipation than the artificial quantum systems, and the DTC in solids has not been fully explored yet.

\begin{figure*}[t]
	\includegraphics[width=\linewidth]{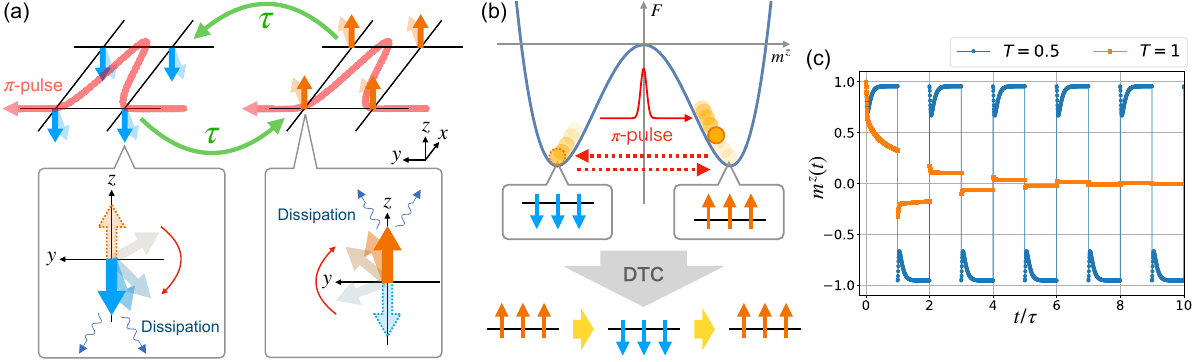}
	\caption{
	(a) Schematic illustration of DTC in our model with $d=2$. Two ferromagnetic states, which the dissipation stabilizes, are switched by $\pi$-pulses.
	(b) Free energy picture for our DTC.
	A state goes back and forth between two minima of the equilibrium free energy by $\pi$-pulses while feeling the past mean-field.
	The dissipation pushes the state into one of the minima, stabilizing the DTC.
	(c) Time profiles of magnetization $m^z(t)$ obtained by the time-dependent mean-field theory for $T=0.5$ (blue circles) and $1$ (orange squares) with $\tau=10$. The initial state is $\rho(0)=\ket{\uparrow}\bra{\uparrow}$. 
	}
	\label{fig:illust}
\end{figure*}

In this paper, we propose a way of realizing a dissipative DTC in solid-state materials subject to generic dissipation to a low-temperature heat bath, investigating it for a concrete model.
For illustration, let us consider a uniaxial ferromagnet as an example,
in which two ferromagnetic states are switched alternately by consecutive $\pi$-pulses, becoming a DTC (see Fig.~\ref{fig:illust}(a)).
This DTC state can exist even in the presence of the dissipation of solids
because the quantum coherence between the two ferromagnetic states is unnecessary for realizing the DTC. 
The dissipation destroys the quantum superposition between them (i.e., the cat state), materializes either ferromagnetic state, and thereby stabilizes the DTC behavior rather than destroys.
In other words, the stability of the ferromagnetic states in solids leads to the rigidity of the DTC~\cite{Gambetta2019a}.

To verify this scenario microscopically, we consider a simple model of the dissipative DTC in solids, a dissipative quantum Ising model periodically driven by a train of $\pi$-pulses.
Describing the microscopic dissipation due to thermal contact to a heat bath like phonons by the Bloch-Redfield (BR) equation~\cite{BreuerBook}, we elucidate that the dissipation stabilizes the DTC, where two ferromagnetic states are switched by each $\pi$-pulse (see Fig.~\ref{fig:illust}(a)).
We also show that the DTC phase transition is continuous and identify its critical behavior using the time-dependent mean-field theory that solves self-consistently the periodic drive, dissipation, and DTC order parameter.
Remarkably, a novel reentrant transition without equilibrium counterparts occurs when changing the pulse interval due to a nontrivial interplay of the periodic drive and dissipation.
Furthermore, we demonstrate the rigidity, or robustness, of the DTC against small imperfections of the $\pi$-pulses and find that
the discrete time quasicrystal (DTQC)~\cite{Giergiel2019b,Zhao2019,Pizzi2019,Chinzei2020} can appear for the larger imperfections.
This rigidity implies that our DTC can be realized in solid-state experiments, where various noises and perturbations are unavoidable.

The structure of the paper is as follows.
In Sec.~\ref{sec:model}, we introduce a theoretical model for the dissipative DTC in solids: a dissipative quantum Ising model driven by a train of $\pi$-pulses, where the BR equation describes the dissipation.
In Sec.~\ref{sec:FMFT}, in order to analyze the spontaneous symmetry breaking accompanied by the DTC phase transition, we invoke the time-dependent mean-field theory and solve the DTC order parameter self-consistently.
In Sec.~\ref{sec:trans}, we numerically and analytically investigate the phase transition and criticality of the DTC based on the mean-field theory.
In Sec.~\ref{sec:rigid}, we demonstrate the rigidity of our DTC against weak perturbation and show an existence of DTQC under strong perturbation. 
In Sec.~\ref{sec:ex}, we discuss the possible experimental realizations of our DTC.
Finally, we summarize this paper and present future works in Sec.~\ref{sec:fin}.

\section{Model of time crystal in solids}\label{sec:model}

\subsection{Quantum Ising model and basic picture}
To investigate dissipative DTCs in solids, we consider a quantum Ising model on the $d$-dimensional square lattice periodically driven by a sequence of $\pi$-pulses with ideal zero pulse width (we analyze the case of finite pulse width in Appendix~\ref{secap:finitew}).
The Hamiltonian is given by
\begin{align}
	H(t) = - J \sum_{\langle i,j\rangle} \sigma^z_i \sigma^z_j + \frac{\pi}{2} \sum_{j,n} \delta(t-n\tau)\sigma^x_j, \label{eq:Ham}
\end{align}
where $\tau$ denotes the time interval between the consecutive $\pi$-pulses serving as the time period
\begin{align}
H(t+\tau)=H(t)
\end{align}
and $J$ is the Ising interaction.
For clarity, we focus on the ferromagnetic interactions $J>0$ and the ferromagnetic DTC.
Yet, the following argument equally applies to the antiferromagnetic ones $J<0$ and the N\'{e}el-like DTC (hence a space-time crystal) since these two cases are mathematically equivalent under the transformation $\sigma_i^z\to-\sigma_i^z$ on either sublattice.

Without dissipation, the Hamiltonian~\eqref{eq:Ham} exhibits the DTC due to Floquet dynamical symmetry~\cite{Else2016,VonKeyserlingk2016,Yao2017,Chinzei2020,Tindall2019,Buca2019,Buca2021, Sarkar2021}
\begin{align}
{U_F}\sigma^z_j {U_F^\dag}=-\sigma^z_j, \label{eq:FDS0}
\end{align}
where $U_F={\mT}e^{-i\int_0^{\tau}dtH(t)}$ is the one-cycle time evolution operator ($\mT$ is the time ordering operator and we set $\hbar=k_B=1$ throughout this paper).
This symmetry leads to the time-crystalline evolution with period $2\tau$,
\begin{align}
\braket{\sigma^z_j(t=t_0+n\tau)} = (-1)^n \braket{\sigma_j^z(t=t_0)},
\end{align}
where $\sigma^z_j(t)$ is in the Heisenberg picture, and $\braket{\cdots}$ denotes the expectation value taken for an arbitrary initial state.
Nevertheless, this time-crystalline nature is fragile against symmetry-breaking perturbations without additional stabilization mechanisms such as many-body localization~\cite{Else2016,VonKeyserlingk2016,Yao2017}. 

One thus might anticipate that dissipation generically breaks the DTC.
However, this is not necessarily true in solid-state materials.
In solids, a typical dissipation is caused by couplings to heat baths at temperature $T$, bringing the system of interest to the same temperature.
Therefore, if $T$ is low enough, the dissipation tends to cool our spin system, and the ferromagnetic order should be favored.
When flipped by the consecutive $\pi$-pulses, this ferromagnetic state would be a DTC as illustrated in Fig.~\ref{fig:illust}(a).
This mechanism can be described by free energy picture as shown in Fig.~\ref{fig:illust}(b), 
where a state goes back and forth between two minima of equilibrium free energy (corresponding to the two ferromagnetic states) by $\pi$-pulses.
This picture lets us imagine the rigidity of the DTC: Even if there exist perturbations disturbing the order, the dissipation brings the state to a nearby free-energy minimum.
Although this argument gives a simple interpretation, we need, for complete understanding, a quantitative microscopic theory, by which we will obtain critical exponents and find richer phenomena such as a nontrivial reentrant DTC transition in varying the pulse interval $\tau$.

\subsection{Bloch-Redfield equation}
To describe dissipation microscopically, we consider a situation that each spin component at all sites is coupled to a huge bosonic heat bath at temperature $T$, where the system-bath coupling is given by 
\begin{align}
	H_\text{SB} = \sum_{j,\mu} \sqrt{\lambda_\mu} \sigma^\mu_j \otimes B^\mu_j.
\end{align}
Here $\lambda_\mu$ is the dimensionless strength of the system-bath coupling for channel $\mu=x,y,z$ and $B^\mu_j$ is an operator for the bath degree of freedom ($B^\mu_j$ has the energy's dimension in our units).
For simplicity, we neglect correlations between different bath degrees of freedom $B^\mu_j$:
$\braket{B^\mu_j(t)B^{\mu'}_{j'}(t')}=\delta_{\mu\mu'}\delta_{jj'}\gamma(t-t')$.
Here, $\gamma(t)$ is the bath correlation function, which we assume ohmic~\cite{Nathan2020},
\begin{align}
	\gamma(t) = \int_{-\infty}^\infty d\epsilon \tgamma(\epsilon) e^{-i\epsilon t}, \,\,\,\,\,\,
	\tgamma(\epsilon) = \frac{\epsilon e^{-\frac{\epsilon^2}{2\Lambda^2}}}{1-e^{-\epsilon/T}},
\end{align}
where $\Lambda$ is the bath spectral cutoff set as $\Lambda=5$ in this work.
We note that $\tgamma(\epsilon)$ satisfies the so-called Kubo-Martin-Schwinger (KMS) condition, 
\begin{align}
	\tgamma(-\epsilon)=e^{-\epsilon/T}\tgamma(\epsilon), \label{eq:KMS}
\end{align}
which leads to the thermal equilibrium of the system without the $\pi$-pulses in the weak coupling limit~\cite{BreuerBook}.

Tracing out the bath degrees of freedom and using the Born-Markov approximation, the density matrix of the system $\rho$ obeys the following BR equation~\cite{BreuerBook},
\begin{align}
	\partial_t \rho &= \mR_t(\rho) \notag \\
	&= -i[H(t),\rho] - \sum_{j,\mu} \lambda_\mu \left( \left[\sigma^\mu_j, \Sigma^\mu_j(t)\rho \right] + \text{h.c.} \right), \label{eq:oriBRE}
\end{align}
with 
\begin{align}
	&\Sigma^\mu_j(t) = \int_{-\infty}^t dt' \gamma(t-t') U(t,t') \sigma^\mu_j U^\dag(t,t'), \\
	&U(t,t')\equiv\mT \exp\left[-i\int_{t'}^t ds H(s)\right].
\end{align}
We note that although the BR equation, in general, can break the positivity of the density matrix, it is not broken  in all the results of this paper.

We set the system-bath coupling to respect the U(1) symmetry (spin rotation around the $z$-axis) of the undriven Hamiltonian $H_0=-J\sum\sigma^z_i \sigma^z_j$.
One can easily verify that, for $\lambda_x = \lambda_y$, the BR equation has the U(1) symmetry: $[\mR_t, \mV_\phi]=0$ ($\mV_\phi(\rho) =  e^{i\phi\sigma^z_\mathrm{tot}}\rho e^{-i\phi\sigma^z_\mathrm{tot}}$ with $\sigma^z_\mathrm{tot}\equiv\sum_i \sigma^z_i$).
In other words, we focus on magnetic materials that are spin-U(1) symmetric, including dissipation.
In Eq.~\eqref{eq:Ham}, we have taken the $\pi$-pulse along the $x$-axis, which could have been any direction in the $x$-$y$ plane. Yet, our choice does not lose generality thanks to the U(1) symmetry.
In the following, we set $\lambda_x=\lambda_y=0.05$ and $\lambda_z=0.1$ unless otherwise mentioned.

\section{Time-dependent mean-field theory} \label{sec:FMFT}

In the presence of dissipation, time-periodic drives usually bring the system, in the long-time limit, to the nonequilibrium steady state (NESS) that oscillates with the same period: $\rho_\mathrm{ness}(t+\tau)=\rho_\mathrm{ness}(t)$.
This means that the discrete time-translation symmetry of the BR equation $\mR_t=\mR_{t+\tau}$ is not broken in $\rho_\mathrm{ness}(t)$.
In fact, the NESS of our model does not break the symmetry for a finite system size.
Here, however, we argue that the many-body interaction between the spins can break this symmetry in the NESS in the thermodynamic limit, giving rise to the DTC.

To analyze the spontaneous breaking of the discrete time translation symmetry, we invoke the time-dependent mean-field theory~\cite{Esin2021}.
Here let us consider a time-dependent order parameter that is uniform in space:
\begin{align}
	m^z(t) = \text{Tr}[\sigma^z_j\rho(t)].
\end{align}
Then, the mean-field Hamiltonian is given by
\begin{align}
	H_\MF(t) = - m^z(t) \sigma^z + \frac{\pi}{2} \sum_{n} \delta(t-n\tau)\sigma^x,
\end{align}
where we have set $Jd=1$ as the unit of energy and omitted the site index $j$ since the mean-field Hamiltonian is decoupled for each site.

In the mean-field approximation, the density matrix $\rho$ obeys the following BR equation for a given mean-field $m^z(t)$: 
\begin{align}
	\partial_t \rho &= -i[H_\MF(t),\rho] \notag \\
	&\hspace{1cm}- \sum_{\mu} \lambda_\mu \left( \left[\sigma^\mu, \Sigma^\mu_\MF(t)\rho \right] + \text{h.c.} \right), \label{eq:mfBRE}
\end{align}
with 
\begin{align}
	&\Sigma^\mu_\MF(t) = \int_{-\infty}^t dt' \gamma(t-t') U_\MF(t,t') \sigma^\mu U^\dag_\MF(t,t'), \\
	&U_\MF(t,t') \equiv \mT \exp\left[-i\int_{t'}^t ds H_\MF(s)\right].
\end{align}
Meanwhile, the mean-field $m^z(t)$ should satisfy the following time evolution equation:
\begin{align}
	\partial_t m^z = \text{Tr}[\sigma^z (\partial_t \rho)]. \label{eq:mfmz}
\end{align}
Therefore, solving Eqs.~\eqref{eq:mfBRE} and \eqref{eq:mfmz} simultaneously,
we obtain the self-consistent solution for $m^z(t)$.
In the numerical calculations, we use the forth-order Runge-Kutta method, solving Eqs.~\eqref{eq:mfBRE} and \eqref{eq:mfmz}.

The mean-field BR equation~\eqref{eq:mfBRE} has nontrivial non-Markovianity, or memory effect,
in the sense that $\Sigma_\MF^\mu(t)$ involves the information of the past state $\rho(t')$ ($t'<t$) via the mean-field $m^z(t') = \text{Tr}[\sigma^z \rho(t')]$.
This non-Markovianity derives from the fact that the system has been evolved under the mean-field Hamiltonian $H_\MF(t')$ involving $m^z(t')$.
Because of this memory effect, the spins at time $t$ are affected by the past state at $t'<t$ with correlation $\gamma(t-t')$ and tend to face the same direction as the past (see also Fig.~\ref{fig:illust}(b)).
We note that the memory time, namely the width of $\gamma(t)$, is finite, which is $O(1/\Lambda)$.
The interplay of the periodic drive and the memory effect stemming from the dissipation gives rise to a rich phase diagram, as shown below.
We remark that this non-Markovianity is different from the Markovian approximation made in deriving the BR equation, which concerns the memory effect of the bath itself.

Let us illustrate the solutions of Eqs.~\eqref{eq:mfBRE} and \eqref{eq:mfmz} in Fig.~\ref{fig:illust}(c), where the time evolution of $m^z(t)$ from an initial state $\rho(0)=\ket{\ua}\bra{\ua}$ is shown for a low and a high temperature.
At the low temperature ($T=0.5$),
$m^z(t)$ relaxes to NESS with $m^z(t)=-m^z(t+\tau)$,
implying the emergence of period doubling $m^z(t+2\tau)=m^z(t)$, i.e., the DTC behavior.
On the other hand,
at the high temperature ($T=1$), $m^z(t)$ decays to vanish after a long time, meaning that the system is in the normal phase where the symmetry is not broken.
The time profile for $T=0.5$ also shows the nontrivial memory effect.
In the NESS, $|m^z(t)|$ decreases just after a pulse is applied at $t=n\tau$ ($n\in\mathbb{Z}$), 
which implies that the spins feel the past mean-field and are about to face the same direction as that in the past due to the memory effect.
We note that, for small $\tau$, the mean-field $m^z(t)$ oscillates very quickly, the memory effect is thereby effectively suppressed, and the dips of $|m^z(t)|$ at $t=n\tau$ become small (see Fig.~\ref{fig:delta}(b) in Appendix~\ref{secap:finitew}).

This DTC transition at lower temperature accompanies a  spontaneous $\mathbb{Z}_2$ symmetry breaking.
Our Hamiltonian originally has the discrete time translation symmetry $\mathbb{Z}$, $H(t) = H(t+n\tau)$ ($n \in \mathbb{Z}$),
and the spin $\pi$-rotation symmetry around the $x$-axis $\mathbb{Z}_2$, $[H(t),P]=0$ ($P=\Pi_j \sigma^x_j$),
which also hold in the BR equation, $\mR_t=\mR_{t+n\tau}$ and $[\mR_t, \mP]=0$ ($\mP(\rho) \equiv P \rho P$).
In the DTC phase, the total symmetry $\mathbb{Z} \times \mathbb{Z}_2$ is broken to $\mathbb{Z}$ due to the many-body interaction,
which is characterized by the following dynamical symmetry~\cite{Alon1998}:
\begin{align}
	H(t+n\tau) = P^n H(t) P^n \qquad (n\in\mathbb{Z}).\label{eq:dynsym}
\end{align}
In other words, the mean-field with $m^z(t)=-m^z(t+\tau)$ emerges in the DTC phase,
which is consistent with Fig.~\ref{fig:illust}(c).
As far as the authors investigated,
the dynamical symmetry~\eqref{eq:dynsym} is not broken for any $T$ and $\tau$.
Thus, our DTC is a $\mathbb{Z}_2$-symmetry broken state of $\mathbb{Z}\times\mathbb{Z}_2 \rightarrow \mathbb{Z}$.

\begin{figure*}[t]
	\includegraphics[width=\linewidth]{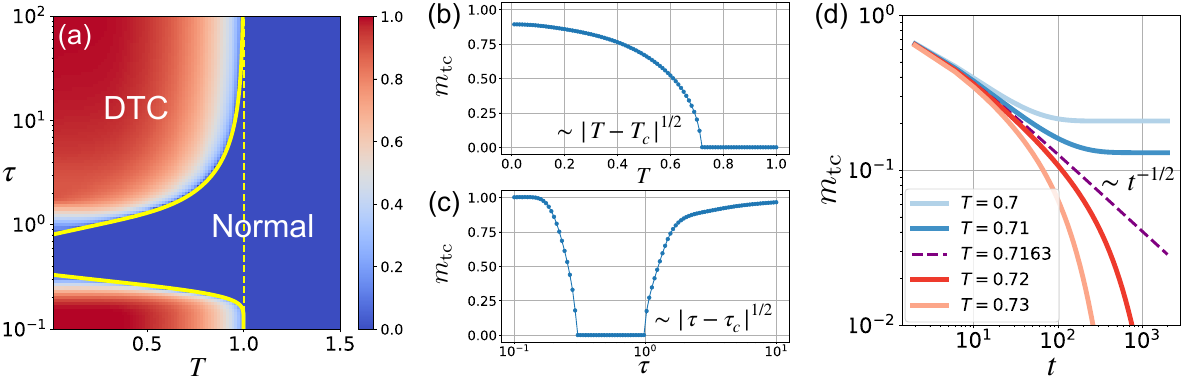}
	\caption{
	(a) Phase diagram for $\mtc$ on $T$-$\tau$ plane. The color denotes the magnitude of $\mtc$ and the red (blue) region corresponds to DTC (normal) phase.
	The yellow solid line is obtained from the exact analysis, $p(T_c,\tau_c)=1$, and the yellow dashed line indicates the equilibrium transition temperature $T=1$.
	(b) $T$-dependence of $\mtc$ for $\tau=2$ and
	(c) $\tau$-dependence of $\mtc$ for $T=0.2$. 
	Their critical exponents are both $1/2$.
	(d) Relaxation dynamics of $\mtc(t_n)$ in the vicinity of critical temperature $T_c \sim 0.7163$ for $\tau=2$. 
	For visibility, $\mtc$ in every time-period $2\tau$ are plotted.
	On the critical point, $\mtc(t_n)$ decays in power-law as $\mtc(t_n)\sim t_n^{-1/2}$. 
	All results in these figures are obtained from the initial state $\rho(0)=\ket{\ua}\bra{\ua}$.
	 }
	\label{fig:4tran}
\end{figure*}

\section{Phase transition and criticality} \label{sec:trans}

\subsection{Reentrant transition} 

Now we investigate the phase transition and the criticality on the $T$-$\tau$ plane.
To quantify the symmetry breaking, we introduce the time-crystalline order parameter at $t_n=2n\tau$ ($n\in\mathbb{Z}$):
\begin{align}
	&\mtc(t_n) = \int_{t_n}^{t_n+\tau} \frac{dt}{2\tau} m^z(t) - \int_{t_n+\tau}^{t_n+2\tau} \frac{dt}{2\tau} m^z(t),
\end{align}
and its long-time limit,
\begin{align}
	\mtc\equiv\lim_{n\rightarrow\infty}\mtc(t_n).
\end{align} 
This order parameter detects the DTC as it vanishes if the discrete time translation symmetry or the spin flip symmetry are not broken (i.e., $m^z(t)=m^z(t+\tau)$ or $m^z(t)=0$).

In the long-time limit, the entire phase diagram for $\mtc$ is shown in Fig.~\ref{fig:4tran}(a) on the $T$-$\tau$ plane.
Interestingly, there are two separated DTC regions, and $\mtc$ exhibits the second-order (continuous) transition at the phase boundaries.
Whereas the phase transition occurs once as $T$ varies with $\tau$ fixed, it does twice as $\tau$ changes for constant $T$ (i.e., reentrance of the DTC phase).
This reentrant transition by $\tau$ is absent in equilibrium and thus an essentially nonequilibrium phenomenon.
Closely looking into the order parameter, we find that the critical exponents for the transition in the $T$ and $\tau$ directions are both 1/2: 
\begin{align}
	\mtc \sim |T-T_c|^{1/2} \,\,\, \text{and} \,\,\, \sim |\tau-\tau_c|^{1/2}, \label{eq:mtc_cri}
\end{align}
respectively, in the vicinity of the phase transition.
These behaviors are shown in Figs.~\ref{fig:4tran}(b) and (c) for representative parameters $\tau=2$ and $T=0.2$.

The breakdown of the DTC for the intermediate $\tau$ is due to a competition between the pulse interval $\tau$ and the other time scales, the memory time $O(1/\Lambda)$ and the inverse of the exchange interaction $O(1/J)$.
To understand this intuitively, let us first consider the two limits of $\tau\rightarrow\infty$ and $0$.
For $\tau \rightarrow \infty$, 
after the system is disturbed by each $\pi$-pulse at $t=n\tau$ ($n\in\mathbb{Z}$), it stays undisturbed longer than the memory time and relaxes back to an equilibrium state before another pulse is applied (see Fig.~\ref{fig:illust}(c)).
In contrast, for $\tau \rightarrow 0$, 
the pulse interval is so short that the system cannot respond to the disturbance originating from the drive and the dissipation, which leads to the relaxation to the equilibrium state as well.
Therefore, in these two limits, the transition temperature and criticality are identical to those in the thermal equilibrium of $H_0=-J\sum \sigma^z_i \sigma^z_j$.
Based on the mean-field theory in equilibrium, the transition occurs at $T_c=1$ in $\tau \rightarrow \infty$ and $0$,
which is consistent with Fig.~\ref{fig:4tran}(a).
However, for the intermediate $\tau$ comparable with $O(1/\Lambda)$ and $O(1/J)$,
the multiple time scales compete with each other, the dynamics is disturbed, and the DTC breaks down (see Appendix~\ref{secap:timescale} for detailed discussions).

\subsection{Exact analysis}

Remarkably, we can analytically obtain these phase boundaries and critical exponents considering the weak coupling limit $\lambda_\mu \rightarrow 0$, where the BR equation is valid.
In this limit, $\partial_t m^z(t) \propto \lambda_\mu\rightarrow0$ except $t=n\tau$, and $m^z(t)$ has the following form,
\begin{align}
m^z(t) = \begin{cases}
	+M(t) & (t_n \le t < t_n+\tau) \\
	-M(t) & (t_n+\tau \le t <t_{n+1}),
\end{cases}
\end{align}
where $M(t)$ is a continuous function slowly varying in $t$.
Thus, the BR equation~\eqref{eq:mfBRE} reduces to the time evolution equation for $M(t)$ (see Appendix~\ref{secap:Mt} for derivation):
\begin{align}
	\partial_t M = \alpha(M) - \beta(M)M. \label{eq:partialM}
\end{align}
Here we have defined 
\begin{align}
	&\alpha(M) = X^-(M) + Y^-(M),\\
	&\beta(M) = X^+(M) + Y^+(M),
\end{align}
and
\begin{align}
	X^{\pm}(M) &= 2\pi \lambda_x \left[ \tgamma(\epsilon^+_0) \pm \tgamma(\epsilon^-_0) \right], \\
	Y^{\pm}(M) &= \frac{8\lambda_y}{\pi} \sum_{k=-\infty}^\infty \frac{\tgamma(\epsilon^+_{2k+1}) \pm \tgamma(\epsilon^-_{2k+1})}{(2k+1)^2},
\end{align}
with $\epsilon^\pm_{k} = k\Omega \pm 2M$ and the DTC frequency $\Omega=2\pi/2\tau$.
Here $\epsilon_k^\pm$ correspond to quasienergy differences between the Floquet states of up and down spins that are dressed by $n\pm k$ and $n$ \textit{photons} (not phonons), respectively, originating from the $\pi$-pulses ($n$ and $k$ are integers).
The contribution with $\tgamma(\epsilon_k^\pm)$ in Eq.~\eqref{eq:partialM} derives from transitions between the Floquet states, in which the excess energy of $k$ photons are compensated by the heat bath~\cite{Ikeda2021}.

\begin{figure}[t]
	\includegraphics[width=7cm]{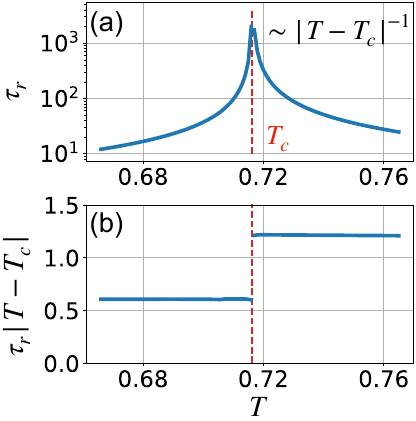}
	\caption{
	Critical behavior of relaxation time $\tau_r$ in Fig.~\ref{fig:4tran}(d).
	The relaxation time diverges at $T=T_c$ as $\tau_r\sim|T-T_c|^{-1}$, and the ratio of coefficients on both sides of the critical point is two.
	 }
	\label{fig:relax}
\end{figure}

\begin{figure}[t]
	\includegraphics[width=7cm]{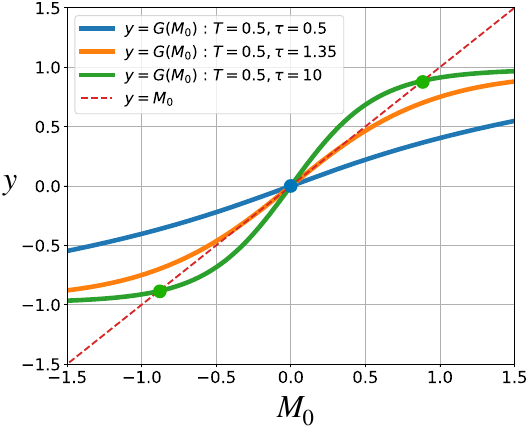}
	\caption{
	Profiles of $G(M_0)$ for $\tau=0.5, 1.35$ and $10$ with $T=0.5$.
	The intersection points of $y=G(M_0)$ and $y=M_0$ correspond to the solutions of Eq.~\eqref{eq:M0SCE}.
	The blue, orange, and green curves denote the normal, critical, and DTC phases, respectively.
	 }
	\label{fig:GM}
\end{figure}

In the long-time limit, $M(t)$ relaxes to a constant $M_0$,
which is determined by the stationary condition $\partial_t M=0$ as
\begin{align}
	M_0=\frac{\alpha(M_0)}{\beta(M_0)} \equiv G(M_0), \label{eq:M0SCE}
\end{align}
where we have used $\beta(M_0)\neq0$.
In the two limits of $\tau\to0$ and $\infty$,
the $k$-photon processes ($k\neq0$) in $G(M_0)$ are negligible because the photon energy exceeds the bath spectral cutoff in $\tau\to0$ and the photon energy vanishes in $\tau\to\infty$.
Therefore, using the KMS condition~\eqref{eq:KMS}, Eq.~\eqref{eq:M0SCE} reduces to the well-known equilibrium self-consistent equation in these limits:
\begin{align}
	M_0=\tanh(M_0/T).
\end{align}
This is why $T_c$ approaches the equilibrium one ($T_c=1$) in Fig.~\ref{fig:4tran}(a) and consistent with the intuition that the memory effect is not relevant when $\tau\to0$ and $\infty$.
In contrast, for the intermediate $\tau$ comparable with the memory time $\sim O(1/\Lambda)$, those $k$-photon processes dominate to disturb the system significantly, destroying the DTC.

The profiles of $G(M_0)$ for representative parameters are depicted in Fig.~\ref{fig:GM},
where $y=M_0$ and $y=G(M_0)$ are plotted, and the intersections between them correspond to the solutions of Eq.~\eqref{eq:M0SCE}. 
Although there is only one intersection at $M_0=0$ in the normal phase (blue curve),
there are three intersections at $M_0=0,\pm m$ in the DTC phase (green curve).
In the DTC phase, the intersections at $M_0=\pm m$ correspond to the $\mathbb{Z}_2$ symmetry-broken state
whereas one at $M_0=0$ is the symmetry-unbroken state that is unstable.
At the critical point, the two curves, $y=M_0$ and $y=G(M_0)$, are tangent at the origin (orange curve).

To obtain the analytic form of the critical point, we expand $G(M_0)$ in Eq.~\eqref{eq:M0SCE} as a power series of $M_0$ since $M_0$ is small near the critical point:
\begin{align}
	G(M_0)=p(T,\tau)M_0 + q(T,\tau)M^3_0 + O(M_0^5), \label{eq:Gexpand}
\end{align}
where $p(T,\tau)$ and $q(T,\tau)$ are functions of $T$ and $\tau$.
We note that there are no even-orders in this expansion because $G(M_0)$ is odd, $G(-M_0)=-G(M_0)$.
Here we assume $q(T,\tau)<0$, which is consistent with Fig.~\ref{fig:GM}.
Using Eq.~\eqref{eq:Gexpand}, the self-consistent equation~\eqref{eq:M0SCE} reduces to
\begin{align}
	M_0 = p(T,\tau)M_0 + q(T,\tau)M^3_0 + O(M_0^5). \label{sm:M3}
\end{align}
The critical point is determined by $p(T_c,\tau_c)=1$.
After a straightforward calculation, we have 
\begin{align}
	p(T,\tau) = \frac{\pi^2 \lambda_x  + 8\lambda_y \sum_k \tgamma'(\epsilon_{2k+1}^0)/(2k+1)^2}{\pi^2 \lambda_x T + 4\lambda_y \sum_k \tgamma(\epsilon_{2k+1}^0)/(2k+1)^2}, \label{eq:yellow}
\end{align}
where we have defined $\epsilon_{k}^0=k\Omega$ and $\tgamma'(\epsilon)=\partial_\epsilon \tgamma(\epsilon)$.
Note that $\tgamma(\epsilon_{2n+1}^0)$ and $\tgamma'(\epsilon_{2n+1}^0)$ depend on $T$ and $\tau$.
The yellow solid curves in Fig.~\ref{fig:4tran}(a) denote the solution of $p(T_c,\tau_c)=1$ for $\lambda_x=\lambda_y$,
which well agrees with the phase boundaries numerically obtained from Eqs.~\eqref{eq:mfBRE} and \eqref{eq:mfmz} for $\lambda_x=\lambda_y=0.05$.

We also obtain the critical behavior~\eqref{eq:mtc_cri} from Eq.~\eqref{sm:M3}.
To this end, we expand $p(T,\tau)$ near the critical point as 
\begin{align}
	p(T,\tau) = 1 + p_1 \delta T + p_2 \delta \tau + O(\delta T^2)+ O(\delta \tau^2), \label{ap:p}
\end{align}
where we have used $p(T_c,\tau_c)=1$ and defined $\delta T=T-T_c$ and $\delta \tau=\tau-\tau_c$ ($p_1$ and $p_2$ are constant depending on $T_c$ and $\tau_c$).
Substituting Eq.~\eqref{ap:p} to Eq.~\eqref{sm:M3} and solving it up to $O(M_0^3)$, 
we obtain the critical behavior of $M_0$ in the lowest order:
\begin{align}
M_0
\begin{cases}
	\sim \big{|} a_1 \delta T + a_2 \delta \tau \big{|}^{1/2}, &\text{for}\hspace{2mm}p(T,\tau)>1\\
	= 0, &\text{for}\hspace{2mm}p(T,\tau)\le1
\end{cases} \label{eq:Msol}
\end{align}
with $a_1=p_1/q_c$ and $a_2=p_2/q_c$ ($q_c=q(T_c,\tau_c)$).
This equation describes the numerically-obtained critical behaviors~\eqref{eq:mtc_cri} in a unified manner.

\begin{figure*}[t]
	\includegraphics[width=\linewidth]{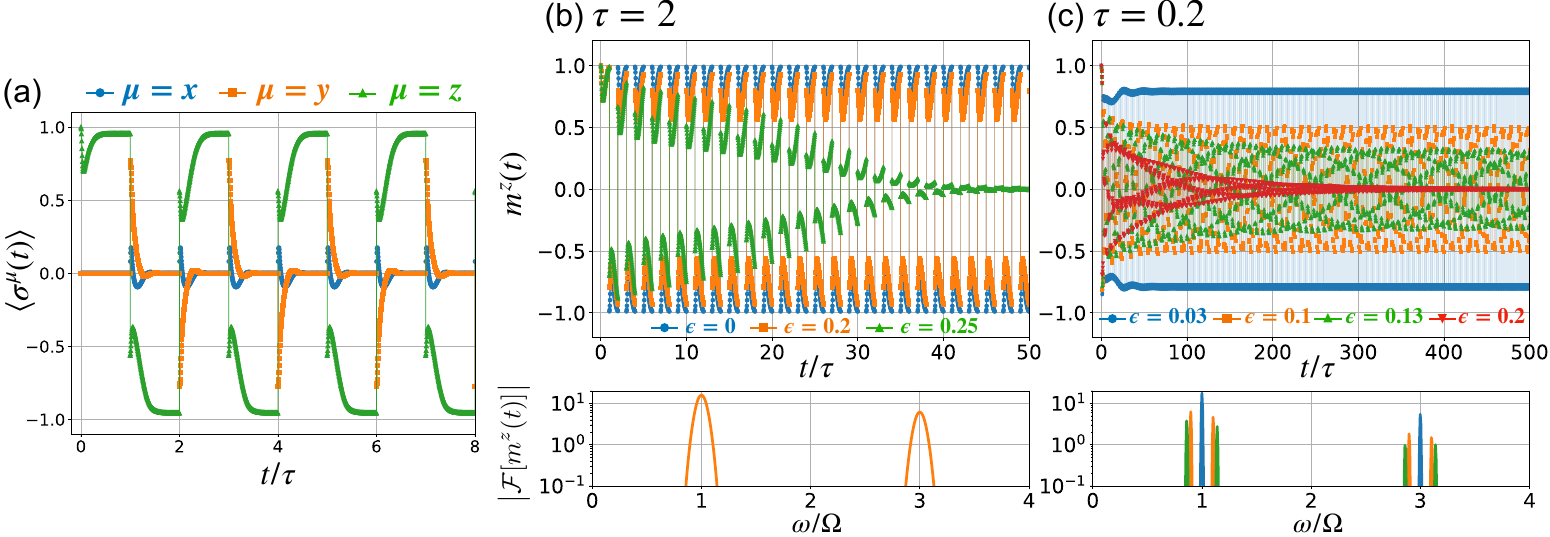}
	\caption{
	(a) Time profiles of $\braket{\sigma^\mu(t)}=\text{Tr}[\sigma^\mu\rho(t)]$ ($\mu=x,y,z$) for $T=0.5, \tau=10$, and $\epsilon=0.3$ ($\braket{\sigma^z(t)}=m^z(t)$).
	Note that, at $t=n\tau$, the pulse only rotates the spin on the $y$-$z$ plane and does not change $\braket{\sigma^x(t)}$ (although it is hard to see because the change after the pulse is quick).
	(b,c) (top) Time profiles of $m^z(t)$ for (a) $\tau=2$ and (b) $\tau=0.2$ with various $\epsilon$'s and (bottom) their Fourier spectra in $t\in[100,200]$, where we have used a window function $w(t)=\exp[-(t-150)^2/20^2]$.
	All results in these figures are obtained from the initial state $\rho(0)=\ket{\ua}\bra{\ua}$.
	 }
	\label{fig:tv_dtqc}
\end{figure*}

\subsection{Dynamical criticality} 

Besides the long-time limit, yet another critical behavior arises during the relaxation dynamics.
Figure~\ref{fig:4tran}(d) shows that, at $T\neq{T_c}$, $\mtc(t_n)$ exponentially relaxes to $\mtc$ with the characterisitic relaxation time $\tau_r$ depending on $T$: $\mtc(t_n) \sim \mtc + (t_n)^{-\chi} \exp(-t_n/\tau_r)$.
At the critical temperature $T=T_c$, $\tau_r$ diverges as (see Fig.~\ref{fig:relax})
\begin{align}
	\tau_r \sim |T-T_c|^{-1}, \label{eq:taur}
\end{align} 
and $\mtc(t_n)$ decays purely in power-law, $\mtc(t_n) \sim (t_n)^{-\chi}$,
where our numerical calculation gives $\chi\sim 1/2$:
\begin{align}
	\mtc(t_n) \sim (t_n)^{-1/2}. \label{eq:mtct}
\end{align} 
Since these critical behaviors are manifest in a finite time window,
they could be directly addressed in experiments (see Sec.~\ref{sec:ex} for further discussion for possible experiments).

We analytically examine the dynamical critical behaviors~\eqref{eq:taur} and \eqref{eq:mtct} from Eq.~\eqref{eq:partialM}.
At the critical point, by expanding the right-hand side of Eq.~\eqref{eq:partialM} up to $O(M^3)$,
we have $\partial_t M \sim q_c \beta(0) M^3$ in the lowest order,
where we have used $G(M) \sim M + q_c M^3$.
Since there are no $O(M)$ terms at the critical point,
we obtain the following power-law decay by solving the differential equation:
\begin{align}
	M(t) \sim t^{-1/2},
\end{align}
which agrees with the numerical result.
Slightly away from the critical point, 
expanding Eq.~\eqref{eq:partialM} by $\delta M(t)=M(t)-M_0$,
we have 
\begin{align}
	\partial_t \delta M \sim \beta(M_0) \left[r(T,\tau)-1 \right] \delta M, \label{eq:deltaM}
\end{align}
where we have defined $r(T,\tau)$ as $G(M) \sim G(M_0) + r(T,\tau)\delta M$.
To obtain $r(T,\tau)$, in Eq.~\eqref{eq:Gexpand}, we expand $G(M)=G(M_0+\delta M)\sim p(T,\tau)(M_0+\delta M) + q(T,\tau)(M_0+\delta M)^3$ in $\delta M$ and compare it with $G(M) \sim G(M_0) + r(T,\tau)\delta M$,
having $r(T,\tau) = p(T,\tau) + 3q(T,\tau) M_0^2$ in the lowest order of $\delta T$ and $\delta \tau$ (note that $M_0^2\sim O(\delta T)$ and $\sim O(\delta \tau)$).
Unlike on the critical point, since Eq.~\eqref{eq:deltaM} has the linear term of $O(\delta M)$,
we obtain the exponential decay:
\begin{align}
	&\delta M(t) \sim \exp\left(-t/\tau_r\right). 
\end{align}
Here $\tau_r$ is the relaxation time
\begin{align}
	&\tau_r \sim
	\begin{dcases}
		\frac{1}{2}\big{|}b_1 \delta T+b_2 \delta\tau\big{|}^{-1}, &\text{for}\hspace{2mm}p(T,\tau)>1 \\
		\big{|}b_1\delta T+b_2\delta\tau\big{|}^{-1}, &\text{for}\hspace{2mm}p(T,\tau)<1
	\end{dcases} 
\end{align}
where we have used $M_0^2\sim[1-p(T,\tau)]/q(T,\tau)$ for the DTC phase and defined $b_1=\beta(M_0)p_1$ and  $b_2=\beta(M_0)p_2$.
This result is consistent with the numerical one, including the ratio between coefficients on both sides of the critical point (see Fig.~\ref{fig:relax}(b)), and it is a generalization of Eq.~\eqref{eq:taur} to the case of $\tau-\tau_c\neq0$.

\section{Rigidity of time crystal} \label{sec:rigid}

As discussed in Sec.~\ref{sec:model}, the free energy picture implies that our time crystal is robust against perturbations, where the dissipation brings the state to a nearby minimum in the free energy and stabilizes the DTC (see Fig.~\ref{fig:illust}(b)).
To verify the rigidity, or robustness against perturbations, of our DTC, we introduce an imperfection $\epsilon$ to the $\pi$-pulse:
 \begin{align}
	H_{\epsilon}(t) = - J \sum_{\langle i,j\rangle} \sigma^z_i \sigma^z_j + \frac{\pi(1+\epsilon)}{2} \sum_{j,n} \delta(t-n\tau)\sigma^x_j. \label{eq:Ham_pe}
\end{align}
Each pulse rotates every spin around the $x$-axis by $\pi(1+\epsilon)$, rather than $\pi$.
Note that $H_{\epsilon}(t)$ has the same symmetry $\mathbb{Z}\times\mathbb{Z}_2$ even for $\epsilon\neq0$.
Without dissipation,
the DTC is broken for $\epsilon\neq0$
since the Floquet dynamical symmetry does not hold, $U_F^\epsilon \sigma^z_j U_F^{\epsilon\dag} \neq -\sigma^z_j$ ($U_F^\epsilon = \mT e^{-i\int_0^\tau dt H_\epsilon(t)}$), unless there exists other stabilizing mechanism such as many-body localization
(see Appendix~\ref{secap:FDS} for the symmetry argument with dissipation).

A representative spin dynamics in the robust dissipative DTC is shown in Fig.~\ref{fig:tv_dtqc}(a).
At $t=n\tau$ $(n\in\mathbb{N})$, each imperfect $\pi$-pulse rotates the spin in the $y$-$z$ plane, instantaneously changing the $y$ and $z$ spin components while keeping the $x$ component unchanged.
The spin then evolves under the dissipation and memory effect, becoming restored to a stable state parallel to the $z$-axis ($\braket{\sigma^{x,y}(t)}=0$ and $m^z(t)=\braket{\sigma^{z}(t)}\sim \pm0.95$), which corresponds to either free-energy minimum.
This observation clearly indicates the origin of rigidity: Even if the pulse is not fine-tuned to the perfect $\pi$-pulse, the dissipation generically pulls the spin back to the $z$-axis and stabilizes the $m^z(t)$'s alternating dynamics, i.e., the DTC order.
We note that the dynamical symmetry~\eqref{eq:dynsym} is not broken in the DTC even for $\epsilon\neq 0$.

The robustness of the DTC against a small imperfection $\epsilon$ holds true for both the large- and small-$\tau$ DTC phases depicted in Fig.~\ref{fig:4tran}(a).
Nonetheless, the large-$\tau$ DTC phase is more stable than that for smaller $\tau$.
Figures~\ref{fig:tv_dtqc}(b) and (c) show $m^z(t)$ for a large-$\tau$ ($\tau=2$) and small-$\tau$ ($\tau=0.2$) for several $\epsilon$'s.
Whereas the DTC is restored up to $\epsilon=0.2$ for $\tau=2$, it is broken already at $\epsilon=0.1$ for $\tau=0.1$.

\begin{figure*}[t]
	\includegraphics[width=\linewidth]{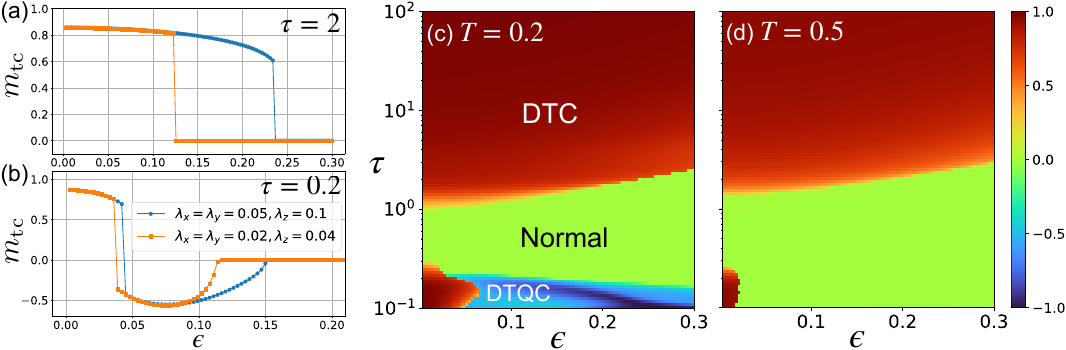}
	\caption{
	(a,b) $\epsilon$-dependence of $\mtc$ for (a) $\tau=2$ and (b) $0.2$ with $(\lambda_x,\lambda_y,\lambda_z)=$  $(0.05,0.05,0.1)$ (blue circles) and $(0.02,0.02,0.04)$ (orange squares).
	The temperature is $T=0.2$,
	and the positive (negative) $\mtc$ denotes the DTC (DTQC) order.
	(c,d) Phase diagrams on $\epsilon$-$\tau$ plane for (c) $T=0.2$ and (d) $0.5$.
	In all panels, $\mtc$ has been obtained in the long-time limit of dynamics starting from the initial state $\rho(0)=0.55 \ket{\ua}\bra{\ua} + 0.45 \ket{\da}\bra{\da}$.
	 }
	\label{fig:eptau}
\end{figure*}

\begin{figure*}[t]
	\includegraphics[width=\linewidth]{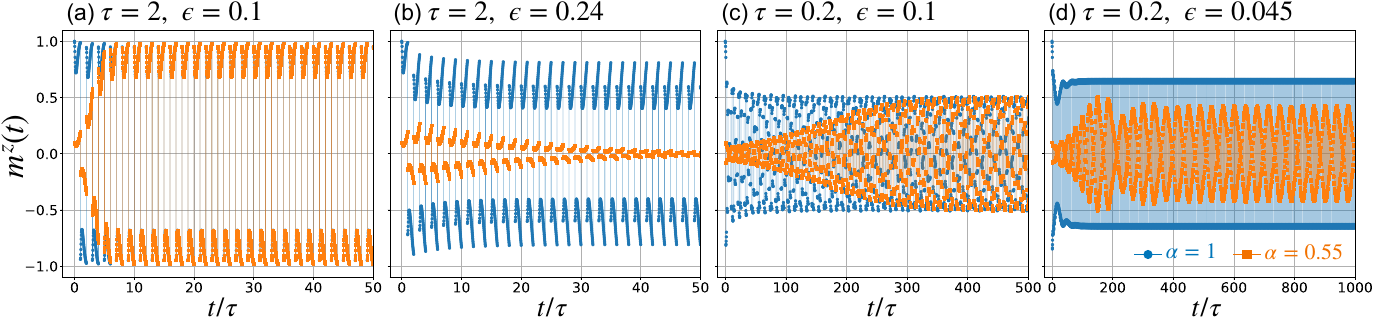}
	\caption{
	(a-d) Time profiles of $m^z(t)$ from initial state $\rho(0)=\alpha\ket{\ua}\bra{\ua} + (1-\alpha)\ket{\da}\bra{\da}$ for $(\tau,\epsilon) = $ (a) $(2,0.1)$, (b) $(2,0.24)$, (c) $(0.2,0.1)$, and (d) $(0.2,0.045)$.
	The blue circles (orange squares) indicate the results for $\alpha=1$ ($0.55$).
	The temperature is $T=0.2$.
	}
	\label{fig:multi}
\end{figure*}

Besides, whereas the DTC directly becomes the normal phase as $\epsilon$ increases for large $\tau$, it is broken first to the DTQC~\cite{Giergiel2019b,Zhao2019,Pizzi2019,Chinzei2020} for small $\tau$, in which $m^z(t)$ never decays but keeps oscillating quasi-periodically, before becoming the normal phase.
This DTQC is seen in Fig.~\ref{fig:tv_dtqc}(c) for the intermediate $\epsilon$ ($=0.1$ and $0.13$) and clearly indicated in the Fourier spectra of $m^z(t)$.
While the DTC is characterized by the Fourier peaks $\omega=(2n+1)\Omega$ (even-order harmonics are absent due to the dynamical symmetry $m^z(t)=-m^z(t+\tau)$ implying $\int_0^{2\tau}dt\, m^z(t)e^{i2n\Omega t}=0$), these peaks are split into two in the DTQC.

Intuitively, these distinct fates of the large- and small-$\tau$ DTC phases against $\epsilon$ originate from how effectively the dissipation stabilizes the DTC. 
As discussed above, the rigidity of DTC stems from the restoring force into the free-energy minima by dissipation.
However, for small-$\tau$ (i.e., short interval), the state cannot be restored before the next pulse arrives,
which leads to the fact that the DTC for small-$\tau$ is more fragile than that for large-$\tau$. 
Yet, even for small $\tau$, the spin does not necessarily decay to vanish because it can be approximately restored being rotated many times by the sequential imperfect pulses for a short time,
which gives rise to the DTQC.

These arguments suggest that the DTC and DTQC become more robust as the dissipation strength $\lambda_\mu$ increases.
In Fig.~\ref{fig:eptau},
we show the $\epsilon$-dependences of $\mtc$ with two dissipation strengths for (a) $\tau=2$ and (b) $0.2$.
Here, for illustration, we have defined $\mtc$ in DTQC as the negative value whose norm is the amplitude of the oscillation of $m^z(t)$ in the NESS.
For $\tau=2$, the transition point from the DTC to the normal phase becomes large as the dissipation strength $\lambda_\mu$ increases.
For $\tau=0.2$, both transition points become large as the dissipation strength while the DTC-DTQC transition point shifts only slightly.
These results show that the DTC and DTQC become more rigid due to the stronger dissipation as expected.

Besides, in Figs.~\ref{fig:eptau}(a) and (b), we can observe the discontinuous, namely first-order, transition from the DTC to the normal (DTQC) phase for $\tau=2$ ($0.2$).
This first-order transition originates from multiple stable states in the vicinity of the transition as shown in Fig.~\ref{fig:multi}, where the time evolutions from different initial states are shown for $\epsilon$'s far from and near the transition points.
While, far from the transition points (Figs.~\ref{fig:multi}(a) and (c)), the different initial states relax to the same DTC or DTQC states,
near the transition points (Figs.~\ref{fig:multi}(b) and (d)), they relax to different states,
which means the existence of multiple stable states in the vicinity of the transition.
These imply that the first-order transition occurs due to a jump from a stable state to another.
We leave, for future work, the systematic study of all possible states appearing for large $\epsilon$ and emphasize again that the DTCs are stable against small $\epsilon$.

We finally show the entire phase diagrams on $\epsilon$-$\tau$ plane for $T=0.2$ and $0.5$ in Figs.~\ref{fig:eptau}(c) and (d),
where we have set the initial state as $\rho(0)=0.55\ket{\ua}\bra{\ua}+0.45\ket{\da}\bra{\da}$.
As shown in these figures, at a high temperature ($T=0.5$), the DTQC phase disappears,
and the area of the DTC region at the bottom becomes smaller than that at a lower temperature ($T=0.2$).
This is because, at higher temperatures, the thermal fluctuation is larger and tends to bring the system to a paramagnetic state with $\braket{\sigma^\mu(t)}=0$.
Therefore, to realize the DTQC or more rigid DTC, the lower temperature is generically more advantageous.

\section{Experimental realizations} \label{sec:ex}

Finally, we argue possible experiments for realizing the dissipative DTC in solids.
While we have assumed the ideal zero pulse width, it is finite in real experiments.
During this pulse width $\delta>0$, the interaction $J$ disturbs the $\pi$-rotation of spins and may destroy the DTC order.
Nevertheless, the mean-field theory suggests that the DTC can appear as long as $\delta$ is small enough (see Appendix~\ref{secap:finitew} for details).
When the width $\delta$ is much smaller than $1/J$, the pulse drive overcomes the disturbance by the interaction and approximately rotates the spins by $\pi$.
After the pulse is switched off, the dissipation brings the system to one of the ferromagnetic states,
stabilizing the DTC order.
It is also necessary for the pulse-free duration $\tau-\delta$ to be sufficiently longer than the relaxation time $1/\lambda_\mu$ for realizing the DTC because, for $\tau-\delta \ll 1/\lambda_\mu$, there is not enough time such that the dissipation stabilizes the order.
We note, however, that these results are based on the mean-field approximation and may quantitatively change when analyzed without the approximation. We leave such beyond-mean-field analyses for future work.

The DTC behavior with finite width pulses was experimentally demonstrated~\cite{Schumacher2003,Shiota2012}, in which intense transverse magnetic field pulses with a short width are periodically applied to magnetic materials, switching their magnetization directions.
Therefore, we think that a dissipative DTC was already realized in this sense, though it has not been inteterpreted so in the DTC research context~\cite{Kebler2021,Ball2021}.
They also support the rigidity of our DTC against noises and imperfections in experiments.

However, these experiments correspond to the large-$\tau$ cases, and the phase transitions and critical phenomena have not been explored yet.
To access these theoretical predictions, one needs ultra-short and -intense pulses at a high repetition rate whose interval $\tau$ is comparable to the exchange interaction time-scale $1/J$ and the memory time $\sim O(1/\Lambda)$ (see below for typical values of $\tau$).
Another possibility is to make use of the electron-spin resonance (ESR) technique~\cite{Kirilyuk2010}. 
In ESR, since the ultra-short laser pulses can be used instead of the transverse field, it is easier to make the setup for short $\tau$.
Since the driving part of the Hamiltonian is replaced with the light-matter interaction in this setup,
it is intriguing to study whether the criticality does not change qualitatively due to the universality.

Table~\ref{tb:unit} provides typical values of the upper and lower critical intervals in the reentrant transition denoted by $\tau_{c1}$ and $\tau_{c2}$, respectively.
Since $\tau_{c1}$ and $\tau_{c2}$ depend on the temperature $T$ (see Fig.~\ref{fig:4tran}(a)), we take two example values $T=0.5$ and $0.9$ as well as two fundamental energy scales $Jd=10$\,meV and $100$\,meV.
This table suggests two ways to obtain larger critical intervals for the upper one $\tau_{c1}$, which are preferable for experimental feasibility.
The first way is to use a magnetic material with small exchange interaction $J$ and bath spectral cutoff $\Lambda$ as $\tau_{c1}$ (and $\tau_{c2}$) is proportional to $1/J$ (note that $\Lambda/Jd$ is fixed).
However, in this approach, since the transition temperature becomes small proportionally to $J$, one must cool the material to lower temperatures, e.g., 105\,K for $T=0.9$ and $Jd=10$\,meV.
The second way is to make the temperature just a little lower than the transition temperature in thermal equilibrium.  
In fact, in the limit of $T \rightarrow T_c^\text{eq} - 0$ ($T_c^\text{eq}$ is the transition temperature in equilibrium), we have $\tau_{c1} \rightarrow \infty$ (see Fig.~\ref{fig:4tran}(a)).
In this approach, while the demand for the small $\tau$ is greatly relaxed, one needs high measurement accuracy to detect the small magnetization as the DTC order parameter is small around $T_c^\text{eq}$.
By appropriately choosing a material and temperature based on these trade-off relations, one could have a chance to detect the transition by $\tau$ and the criticality within the current technologies.

\begin{table}[t]
\caption{
Table of values with units for upper and lower critical intervals $\tau_{c1}$ and $\tau_{c2}$ for $T=0.5$ and $0.9$ depending
on two choices of $Jd=10$\,meV and $100$\,meV.
We have obtained the values from the analytical result~\eqref{eq:yellow} with $\Lambda=5Jd$.
}
  \begin{tabular*}{\linewidth}{@{\extracolsep{\fill}}c|ccc} 
  \hline
  & Energy scale $Jd$  & 10meV & 100meV \\
  \hline \hline
  $T=0.5$\,\, & Temperature $T$  & 58\,K & 580\,K \\
  \,\, & Upper critical point $\tau_{c1}$  & 86\,fs & 8.6\,fs \\
   & Lower critical point $\tau_{c2}$  & 17\,fs & 1.7\,fs \\
  \hline
  $T=0.9$\,\, & Temperature $T$  & 105\,K & 1050\,K \\
  \,\, & Upper critical point $\tau_{c1}$  & 307\,fs & 30.7\,fs \\
  & Lower critical point $\tau_{c2}$  & 13\,fs & 1.3\,fs
  \end{tabular*}
  \label{tb:unit}
\end{table}

\section{Discussions and conclusions} \label{sec:fin}

In this paper, using the time-dependent mean-field theory, 
we have shown that the dissipative time crystals can be realized in solid-state materials and elucidated the criticality and rigidity of them.
While generic dissipation has been expected to destroy time-crystalline behaviors, it rather stabilizes the DTC in our scenario without fine-tuning as long as the temperature is low enough.
Microscopically analyzing our model, we have found the nontrivial transition behaviors without equilibrium counterparts such as
the reentrant transition by changing the pulse interval, which arises from the interplay of the periodic drive and dissipation.
Also, to demonstrate the rigidity,
we have considered the imperfect spin-rotation angle $\epsilon\pi$ away from $\pi$ of each pulse,
showing that the DTC is robust against small $\epsilon$ and finding that the DTQC can appear for large $\epsilon$.
Finally, we have discussed the experimental realizations of our DTC.

We make two remarks on the validity of the mean-field approximation.
First, our mean-field-theory analysis possibly underestimates the Floquet heating,
one of the most significant barriers for realizing the DTC in isolated systems.
It is known that, in generic many-body systems, periodic drives heat the system up to a featureless infinite temperature state~\cite{DAlessio2014,Lazarides2014,Kim2014}.
Our mean-field approximation neglects the many-body correlations, and the heating effect is not evaluated appropriately.
Nevertheless, it is natural to expect that the DTC is indeed realized when energy dissipation rate exceeds the heating rate.
When the pulse interval $\tau$ is longer than the typical relaxation time due to the dissipation $1/\lambda_\mu$, 
the dissipation cools the system faster than the Floquet heating by the drive, stabilizing the DTC.
Also, for small $\tau$ (i.e., high-frequency regime), since it is known that the heating rate is exponentially small in $1/\tau$~\cite{Abanin2015,Kuwahara2016}, even weak dissipation could compensate the heating
(see Appendix~\ref{secap:tau0} for detailed discussions).
To verify whether these expectations are true and determine a more accurate phase diagram, it is necessary to take account of the many-body correlations beyond the mean-field theory. 
Second, the symmetry breaking and the mean-field theory are not entirely formulated in the Floquet dissipative systems.
In this work, since our DTC relies on the equilibrium free energy picture (see Fig.~\ref{fig:illust}(b)), we believe that the mean-field approximation is qualitatively true like in the equilibrium theory, except for the heating problem.
However, the symmetry breaking in the Floquet dissipative systems is not fully elucidated, and one needs further studies to verify the validity of the approximation.

We also note that the BR equation does not guarantee the positivity of the density matrix in general.
While all the results in this paper do not break the positivity, we have encountered the positivity breaking in the several cases of finite pulse width $\delta$.
To analyze such a situation, one should use another theory for open quantum systems ensuring the positivity, such as the Lindblad master equation.

Although we have focused on the $\mathbb{Z}_2$-symmetric materials in this work,
our theory could be extended to $\mathbb{Z}_N$-symmetric ones,
in which $N$ symmetry broken states are switched one after another in every cycle by appropriate pulses, and the DTC with period $N\tau$ is realized.
This could offer a new possibility to create various DTCs in materials,
which is also important from the viewpoint of Floquet engineering in solid-state physics~\cite{Bukov2015,Oka2019}.

Nonequilibrium universality is a crucial open issue.
In this work, we have only focused on the quantum Ising model and the U(1)-symmetric dissipation.
According to the equilibrium theory, the criticality only relies on the symmetries and dimensions, which is known as the universality.
If this holds true for DTCs, the critical exponents that we have found should be common with any models with the Ising symmetry within the mean-field approximation.
Also, finding criticality with different symmetries and calculating critical exponents beyond the mean-field theory are intriguing future directions.
Such a study has recently been reported in the three-dimensional classical Ising model~\cite{Yue2021a}.
We leave further investigations of these questions for future work.

\section*{Acknowledgements}
Fruitful discussions with Akihiko Ikeda and Hirokazu Tsunetsugu are gratefully acknowledged.
K.C. was supported by JSPS KAKENHI Grant No.~21J11245 and Advanced Leading Graduate Course for Photon Science
at the University of Tokyo.
T.N.I. was supported by JSPS KAKENHI Grant No.~JP21K13852.
The computation in this work has been done using the facilities of
the Supercomputer Center, the Institute for Solid State Physics, the
University of Tokyo.

\begin{widetext}
\appendix

\section{Results for finite pulse width} \label{secap:finitew}

\begin{figure*}[t]
	\includegraphics[width=\linewidth]{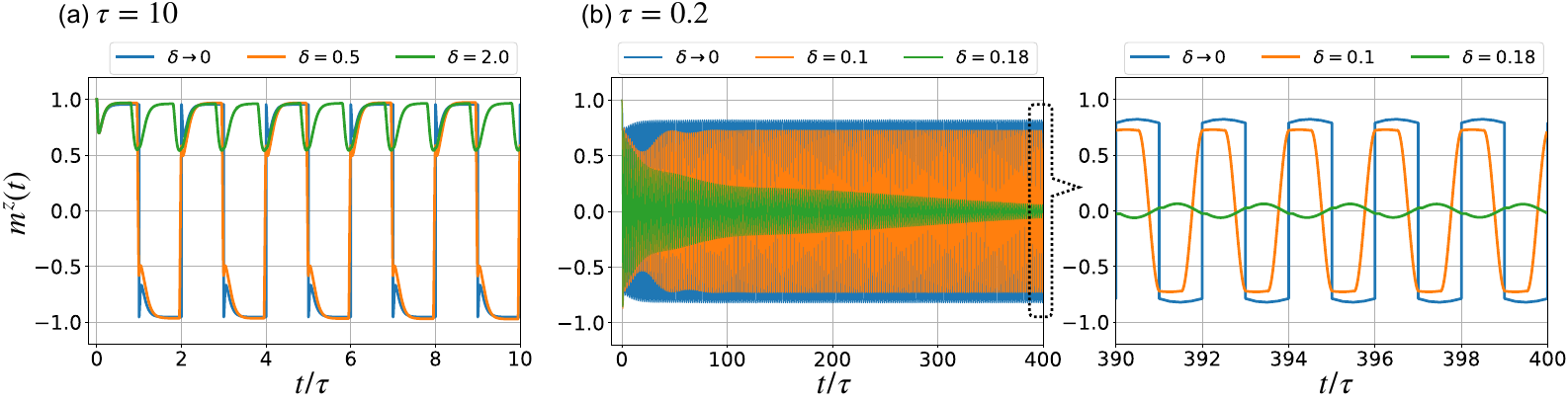}
	\caption{
	(a,b) Time profiles of $m^z(t)$ with various $\delta$ for (a) $\tau=10$ and (b) $\tau=0.2$.
	The right figure in (b) is a magnified view of the left one. 
	The curves with $\delta\to0$ denote the results for the ideal $H(t)$ in the main text.
	The initial state is $\rho(0)=\ket{\ua}\bra{\ua}$, and the temperature is $T=0.5$.
	}
	\label{fig:delta}
\end{figure*}

While we have focused on the ideal pulse of zero width in the main text, 
we examine the case of finite pulse width $\delta > 0$ and show that the DTC survives if $\delta$ is small enough in this Appendix.
Here, let us consider a periodic binary Hamiltonian $H_\text{fw}(t)=H_\text{fw}(t+\tau)$ as follows (we assume $\delta\le\tau$):
\begin{align}
	H_\text{fw}(t) 
	=\begin{dcases}
		H_1=- J \sum_{\langle i,j\rangle} \sigma^z_i \sigma^z_j, & 0 \leq t < \tau-\delta,\\
		H_2=- J \sum_{\langle i,j\rangle} \sigma^z_i \sigma^z_j + \frac{\pi}{2\delta} \sum_{j,n} \sigma^x_j, & \tau-\delta \leq t < \tau.
	\end{dcases} \label{eq:Ham_fw}
\end{align}
The second term in the bottom of Eq.~\eqref{eq:Ham_fw} is a static field along $x$-axis for the finite time window $\delta$.
As $\delta$ decreases, the field amplitude $\pi/(2\delta)$ increases so that each pulse rotates the spins by $\delta\times (\pi/\delta)=\pi$ if $J=0$.
The limit of $\delta\to0$ corresponds to the ideal case of zero pulse width examined in the main text, and thus let $\delta=0$ denote it for convenience.
Unlike in the limit,
the interaction ($J\neq0$) disturbs the rotation of the spins during pulses and may destroy the DTC order.
Here we numerically show that the DTC order actually survives for sufficiently small $\delta$ within the mean-field approximation.
To this end, we solve the BR equation~\eqref{eq:oriBRE} with $H_\text{fw}(t)$ by the forth-order Runge-Kutta method.

Figure~\ref{fig:delta} shows the time evolutions of $m^z(t)$ for various $\delta$.
For (a) $\tau=10$, $m^z(t)$ exhibits the DTC behavior for $\delta=0$ and $=0.5$ whereas does not for $\delta=2.0$,
which indicates that the DTC order is robust against small $\delta$.
This is because, when $\delta$ is much smaller than $1/J$, the strong static field ($\sim1/\delta$) overcomes the disturbance due to the interaction and approximately rotates the spins by $\pi$.
After the static field is switched off, the dissipation brings the state to one of the ferromagnetic states,
stabilizing the DTC order.
On the other hand, when $\delta$ is much larger than $J$, $H_2$ itself exhibits a ferromagnetic order, and the static field cannot rotate the spins sufficiently.
This means that the DTC order is destroyed for $\delta \gg 1/J$.
For (b) $\tau=0.2$, we also observe the DTC behaviors for $\delta=0$ and $=0.1$ whereas does not for $\delta=0.18$.
Note that $\delta$ is much smaller than $1/J$ even for $\delta=0.18$.
This implies that it is necessary for $\delta$ to be sufficiently smaller than not only $1/J$ but also $\tau$ for realizing the DTC.
This is because, for $\delta\sim\tau$, there is not enough time for the dissipation to stabilize the DTC order.
This implies that the duration $\tau-\delta$ should be large enough compared with the relaxation time scale $1/\lambda_\mu$.
As $\lambda_\mu$ increases, the DTC becomes more stable because of the stronger restoring force by the dissipation.

\begin{figure*}[t]
	\includegraphics[width=0.8\linewidth]{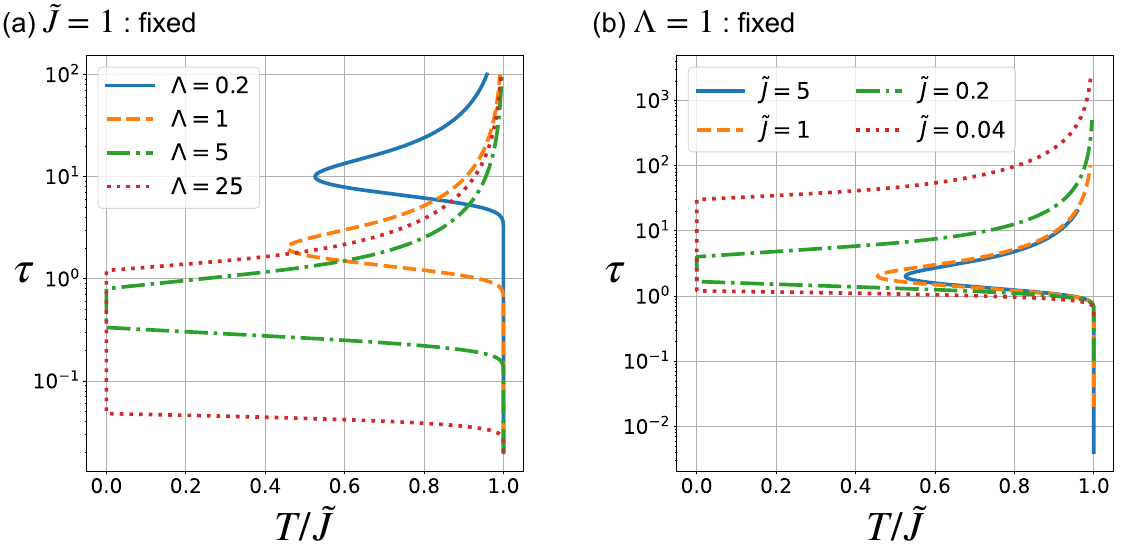}
	\caption{
	(a) Phase boundaries for (solid) $\Lambda=0.2$, (dashed) $\Lambda=1$, (dash-dotted) $\Lambda=5$, and (dotted) $\Lambda=25$ with fixed $\tilde{J}=1$.
	(b) Phase boundaries for (solid) $\tilde{J}=5$, (dashed) $\tilde{J}=1$, (dash-dotted) $\tilde{J}=0.2$, and (dotted) $\tilde{J}=0.04$ with fixed $\Lambda=1$.
	We have obtained these results from the exact solution~\eqref{eq:yellow} for imperfection $\epsilon=0$ and pulse width $\delta\to0$.
	}
	\label{fig:LambdaJ}
\end{figure*}

\section{Examination of phase diagram} \label{secap:timescale}

In this Appendix, we numerically investigate the phase diagrams for the ideal $\pi$-pulse case (i.e., $\epsilon=0$ and $\delta=0$). 
For simplicity, we use the exact result~\eqref{eq:yellow}, $p(T_c,\tau_c)=1$, for the weak coupling limit $\lambda_\mu\to0$.
In the mean-field theory, our model has four independent parameters, $\tau, T, \Lambda$, and $\tilde{J}=zJ$ except $\lambda_\mu$.
While we have chosen $\tilde{J}$ as the unit of energy in the main text,
we will also look into $\tilde{J}$-dependence in this Appendix.

Figure~\ref{fig:LambdaJ} shows the phase boundaries on $T$-$\tau$ plane with various $\tilde{J}$ and $\Lambda$.
In Fig.~\ref{fig:LambdaJ}(a), the results with fixed $\tilde{J}$ and varied $\Lambda$ are shown.
We first notice that the positions of the lower phase boundaries are almost proportional to $1/\Lambda$.
This is consistent with the argument in the main text, in which the ratio of the pulse interval $\tau$ and the memory time (bath correlation time) $1/\Lambda$ determines the phase transition.
On the other hand, the $\Lambda$-dependence of the upper phase boundaries is more complicated.
This stems from the effect of another energy scale, $\tilde{J}$, as shown below.

Figure~\ref{fig:LambdaJ}(b) displays the results with fixed $\Lambda$ and varied $\tilde{J}$.
In the figure, the positions of the lower boundaries hardly depend on $\tilde{J}$, which implies that the lower boundary is approximately determined only by $\Lambda$.
However, the positions of the upper boundaries depend on $\tilde{J}$ (although they are almost the same for $\tilde{J}=1$ and $5$).
In other words, the upper boundary is determined by both $\Lambda$ and $\tilde{J}$, unlike the lower one.
These results highlight that the DTC transition occurs due to the complex interplay of the interaction, the periodic drive, and the dissipation.

\section{Exact analysis and criticality} \label{secap:Mt}

\subsection{Derivation of time evolution equation for $M(t)$} \label{subsecap:Mt}

Here we derive a time evolution equation for $M(t)$ that holds for $\lambda_\mu$ of arbitrary strength.
In Appendix~\ref{subsecap:lambda0}, we will show that this equation reduces to Eq.~\eqref{eq:partialM} in the main text in the weak coupling limit $\lambda_\mu\to0$.

Let us recall the mean-field BR equation:
\begin{align}
	\partial_t \rho &= -i[H_\MF(t),\rho]
	- \sum_{\mu} \lambda_\mu \left( \left[\sigma^\mu, \Sigma^\mu_\MF(t)\rho \right] + \text{h.c.} \right), \label{sm:BReq}
\end{align}
with 
\begin{align}
	&H_\MF(t) = H_0(t) + H_1(t) =- m^z(t) \sigma^z + \frac{\pi}{2} \sum_{n} \delta(t-n\tau)\sigma^x,  \\
	&\Sigma^\mu_\MF(t) = \int_{-\infty}^t dt' \gamma(t-t') U_\MF(t,t') \sigma^\mu U^\dag_\MF(t,t'), \\
	&U_\MF(t,t') \equiv \mT \exp\left[-i\int_{t'}^t ds H_\MF(s)\right],
\end{align}
where we have defined $H_0(t)=-m^z(t) \sigma^z$ and $H_1(t)=(\pi/2)\sum_{n} \delta(t-n\tau)\sigma^x$.
Note that $\Sigma^\mu_\MF(t)$ is non-Hermitian as $\gamma(t-t')$ is complex.
Since each $\pi$-pulse changes the sign of the mean-field $m^z(t)=\Tr[\sigma^z\rho(t)]$ at $t=n\tau$,
we can define a continuous function $M(t)$ as follows,
\begin{align}
M(t) &= v(t)m^z(t), \\
v(t) &= \begin{cases}
	+1 & (t_n \le t < t_n+\tau) \\
	-1 & (t_n+\tau \le t <t_{n+1}),
\end{cases} \label{sm:defv}
\end{align}
with $t_n=2n\tau$ ($n\in\mathbb{Z}$).
Here we derive the time evolution equation for $M(t)$.

For convenience, let us move on to the interaction picture for $H_1(t) = (\pi/2) \sum_{n} \delta(t-n\tau)\sigma^x$:
\begin{align}
	\rho_I(t) = V^\dag(t) \rho(t) V(t),
\end{align}
where we have defined $V(t)=\mT \exp[-i\int_0^t ds H_1(s)]$.
In this picture,
we can rewrite the BR equation~\eqref{sm:BReq} as
\begin{align}
	\partial_t \rho_I &= -i[H_{0,I}(t),\rho_I]
	- \sum_{\mu} \lambda_\mu \left( \left[\sigma^\mu_I(t), \Sigma^\mu_I(t)\rho_I \right] + \text{h.c.} \right), \label{sm:BReqI}
\end{align}
with 
\begin{align}
	H_{0,I}(t) &\equiv V^\dag(t) H_0(t)V(t), \\		
	\sigma^\mu_I(t) &\equiv V^\dag(t) \sigma^\mu V(t), \\
	\Sigma^\mu_I(t) &\equiv V^\dag(t) \Sigma^\mu_\MF(t) V(t)= \int_{-\infty}^t dt' \gamma(t-t') V^\dag(t) U_\MF(t,t') \sigma^\mu U^\dag_\MF(t,t') V(t).
\end{align}
One can easily show that 
\begin{align}
	&\sigma^x_I(t) = \sigma^x, \label{sm:sigmax} \\
	&\sigma^y_I(t) = v(t) \sigma^y, \label{sm:sigmay}\\
	&\sigma^z_I(t) = v(t) \sigma^z \label{sm:sigmaz}
\end{align}
and hence that
\begin{align}
	H_{0,I}(t) = -m^z(t)\sigma^z_I(t) = -m^z(t)v(t)\sigma^z = -M(t)\sigma^z.\label{sm:H0I}
\end{align}
Multiplying $\sigma^z$ and taking the trace of both sides of Eq.~\eqref{sm:BReqI},
we have 
\begin{align}
	\partial_t M(t) = - \sum_{\mu=x,y} \lambda_\mu \Tr  \biggl[ \Bigl( \left[ \sigma^z,\sigma^\mu_I(t) \right]\Sigma_I^\mu(t) + \text{h.c.} \Bigr) \rho_I  \biggr], \label{sm:partialM}
\end{align}
where we have used
\begin{align}
\Tr[\sigma^z \rho_I(t)]=\Tr[V(t)\sigma^z V^\dag(t)\rho(t)]=\Tr[v(t)\sigma^z \rho(t)]=v(t)m^z(t)=M(t) 
\end{align}
on the left-hand side and $\Tr(\sigma^z[H_{0,I}(t),\rho_I])=\Tr([\sigma^z,H_{0,I}(t)]\rho_I)=0$ due to Eq.~\eqref{sm:H0I} on the right-hand side.
We note that the contribution of $\mu=z$ has vanished in Eq.~\eqref{sm:partialM} because $[\sigma^z,\sigma^z_I(t)]=0$ due to Eq.~\eqref{sm:sigmaz}.

To simplify Eq.~\eqref{sm:partialM},
let us look into $\Sigma_I^\mu(t)=\int_{-\infty}^t dt' \gamma(t-t') V^\dag(t) U_\MF(t,t') \sigma^\mu U^\dag_\MF(t,t') V(t)$. 
Here, since the time evolution operator in the interaction picture is given by $U_{\MF,I}(t,t')=\mT\exp[-i\int_{t'}^t ds H_{0,I}(s)] = \exp[i\sigma^z\int_{t'}^t dsM(s)]$,
we can rewrite $U_\MF(t,t')$ as follows
\begin{align}
	U_\MF(t,t') = V(t) U_{\MF,I}(t,t') V^\dag(t') = V(t) e^{i\sigma^z u(t,t')} V^\dag(t'),
\end{align}
where we have defined 
\begin{align}
	u(t,t')=\int_{t'}^t ds M(s).
\end{align}
Therefore, $S_\mu(t,t') \equiv  V^\dag(t) U_\MF(t,t') \sigma^\mu U^\dag_\MF(t,t') V(t)$ in $\Sigma_I^\mu(t)$ reads
\begin{align}
	S_\mu(t,t') 
	&= e^{i\sigma^z u(t,t')} V^\dag(t') \sigma^\mu V(t') e^{-i\sigma^z u(t,t')} \notag \\
	&= e^{i\sigma^z u(t,t')} \sigma_I^\mu(t') e^{-i\sigma^z u(t,t')}, \label{sm:Smu}
\end{align}
and, using Eqs.~\eqref{sm:sigmax} and \eqref{sm:sigmay} together with Eq.~\eqref{sm:Smu}, we have
\begin{align}
	&S_x(t,t') 
	=
	\begin{pmatrix}
	0 & e^{2iu(t,t')} \\
	e^{-2iu(t,t')} & 0 \\
	\end{pmatrix}, \\
	&S_y(t,t') 
	=
	\begin{pmatrix}
	0 & -iv(t')e^{2iu(t,t')} \\
	iv(t')e^{-2iu(t,t')} & 0 \\
	\end{pmatrix}.
\end{align}
By introducing 
\begin{align}
	&a_x^{\pm}(t) \equiv \int_{-\infty}^t dt' \gamma(t-t') e^{\pm2iu(t,t')} = \int_0^\infty ds \gamma(s) e^{\pm2iu(t,t-s)}, \\
	&a_y^{\pm}(t) \equiv \int_{-\infty}^t dt' \gamma(t-t') v(t') e^{\pm2iu(t,t')} = \int_0^\infty ds \gamma(s) v(t-s) e^{\pm2iu(t,t-s)},
\end{align}
we can calculate $\Sigma_I^\mu(t)=\int_{-\infty}^t dt' \gamma(t-t') S_\mu(t,t')$:
\begin{align}
	&\Sigma^x_I(t) 
	=
	\begin{pmatrix}
	0 & a_x^+(t) \\
	a_x^-(t) & 0 \\
	\end{pmatrix}
	= \frac{a_x^-(t)+a_x^+(t)}{2}\sigma^x + \frac{a_x^-(t)-a_x^+(t)}{2i}\sigma^y, \label{sm:Sigmax}\\
	&\Sigma^y_I(t) 
	=
	\begin{pmatrix}
	0 & -ia_y^+(t) \\
	ia_y^-(t) & 0 \\
	\end{pmatrix}
	= \frac{a_y^-(t)-a_y^+(t)}{-2i}\sigma^x + \frac{a_y^-(t)+a_y^+(t)}{2}\sigma^y. \label{sm:Sigmay}
\end{align}
Finally, substituting Eqs.~\eqref{sm:sigmax}, \eqref{sm:sigmay}, \eqref{sm:Sigmax} and \eqref{sm:Sigmay} into Eq.~\eqref{sm:partialM},
we obtain the time evolution equation for $M(t)$,
\begin{align}
	\partial_t M(t) = \xi(t) - \eta(t)M(t), \label{sm:partialM2}
\end{align}
where we have defined 
\begin{align}
	\xi(t)
	&= \lambda_x \bigl[ a_x^+(t) - a_x^-(t) + \text{c.c}\bigr] + \lambda_y v(t) \bigl[ a_y^+(t)-a_y^-(t) + \text{c.c}\bigr], \\
	\eta(t)
	&= \lambda_x \bigl[ a_x^+(t)+a_x^-(t) + \text{c.c}\bigr] + \lambda_y v(t) \bigl[ a_y^+(t)+a_y^-(t) + \text{c.c}\bigr],
\end{align}
and used $\Tr[\rho_I(t)]=1$.
Note that both $\xi(t)$ and $\eta(t)$ are functionals of $M(t)$.
For convenience, we introduce
\begin{align}
	&b_x^\pm(t) = \left[ a_x^\pm(t) + (a_x^\pm(t))^\ast \right] = \int_0^\infty ds \gamma(s) e^{\pm2iu(t,t-s)} + \int_{-\infty}^0 ds \gamma(s) e^{\mp2iu(t,t+s)}, \label{sm:bxpm}\\
	&b_y^\pm(t) = v(t) \left[ a_y^\pm(t) + (a_y^\pm(t))^\ast \right] = \int_0^\infty ds \gamma(s)w(t,s) e^{\pm2iu(t,t-s)} + \int_{-\infty}^0 ds \gamma(s)w(t,s) e^{\mp2iu(t,t+s)},\label{sm:bypm}
\end{align}
where we have used $\gamma(s)=\gamma(-s)^\ast$ and defined
\begin{align}
w(t,s) = \begin{cases}
	v(t)v(t-s) & (s>0) \\
	v(t)v(t+s) & (s<0).
\end{cases}
\end{align}
Then $\xi(t)$ and $\eta(t)$ are written as
\begin{align}
	\xi(t)
	&= \sum_{\mu=x,y} \lambda_\mu \left[ b_\mu^+(t) - b_\mu^-(t) \right], \\
	\eta(t)
	&= \sum_{\mu=x,y} \lambda_\mu \left[ b_\mu^+(t) + b_\mu^-(t) \right].
\end{align}

\subsection{Analytic form for $\lambda_\mu\to0$}\label{subsecap:lambda0}

Here we take the limit of $\lambda_\mu\to0$ and show that Eq.~\eqref{sm:partialM2} reduces to Eq.~\eqref{eq:partialM} in the main text.
To this end, 
we invoke the following two approximations (i) and (ii), which are justified in this limit.

\vspace{5mm}
\noindent
(i)
In the limit of $\lambda_\mu\to0$, $M(t)$ varies very slowly in time ($\partial_t M(t) \sim O(\lambda_\mu)\to0$), and
we can approximate $u(t,t\pm s)=\int_{t\pm s}^t dt' M(t')$ in Eqs.~\eqref{sm:bxpm} and \eqref{sm:bypm} as 
\begin{align}
	u(t,t\pm s)  \sim \pm M(t)s. \label{sm:u2}
\end{align}
Note that $\gamma(s)$ has a finite memory time of $O(\Lambda^{-1})$ and $|\gamma(s)|$ rapidly decays for $|s|\gtrsim\Lambda^{-1}$.
Thus, the semi-infinite integrals over $s$ in Eqs.~\eqref{sm:bxpm} and \eqref{sm:bypm} are actually dominated by $|s|\lesssim\Lambda^{-1}$.
Therefore, the approximation~\eqref{sm:u2} is justified in Eqs.~\eqref{sm:bxpm} and \eqref{sm:bypm} if $\lambda_\mu$ is so small that $\lambda_\mu/\Lambda\ll1$.

Using Eq.~\eqref{sm:u2}, we have
\begin{align}
	&b_x^\pm(t) \sim \int_{-\infty}^\infty ds \gamma(s) e^{\pm 2iM(t)s} = \sum_{k=-\infty}^\infty e^{-2ik\Omega t} c_{x}^\pm(k,t), \\
	&b_y^\pm(t) \sim \int_{-\infty}^\infty ds \gamma(s) w(t,s) e^{\pm 2iM(t)s} = \sum_{k=-\infty}^\infty e^{-2ik\Omega t} c_{y}^\pm(k,t),
\end{align}
where we have defined
\begin{align}
	&c_{x}^\pm(k,t) = \delta_{k0} 2\pi \tgamma(\epsilon_0^\pm(t)), \\
	&c_{y}^\pm(k,t) = \left(\frac{2}{\pi}\right)^2 \sum_{\ell=-\infty}^{\infty} \int_{-\infty}^\infty ds \frac{\gamma(s)e^{\pm2iM(t)s}e^{i(2\ell+1)\Omega s}}{(2\ell+1)(2\ell+1-\text{sgn}(s)k)}. \label{sm:cykt}
\end{align}
To derive these, we have used $2\pi\tgamma(\omega)=\int_{-\infty}^\infty ds\gamma(s)e^{i\omega s}$ and $v(t) = (2i/\pi)\sum_k e^{-i(2k+1)\Omega t}/(2k+1)$.
Here, the time-dependences of $c_\mu^\pm(k,t)$ stem from $M(t)$, and, therefore, $c_\mu^\pm(k,t)$ are slowly varying functions in $t$ ($\partial_t c_\mu^\pm(k,t) \sim O(\lambda_\mu)$).
In particular, the dc components ($k=0$), $c_x^\pm(0,t)$ and $c_y^\pm(0,t)$, are given by
\begin{align}
	&c_{x}^\pm(0,t) = 2\pi \tgamma(\epsilon_0^\pm(t)), \\
	&c_{y}^\pm(0,t) = \frac{8}{\pi} \sum_{\ell=-\infty}^{\infty} \frac{\tgamma(\epsilon_{2\ell+1}^\pm(t))}{(2\ell+1)^2}, 
\end{align}
with $\epsilon^\pm_k(t) = k\Omega \pm 2M(t)$.

\vspace{5mm}
\noindent
(ii)
As the second approximation, in Eq.~\eqref{sm:partialM2}, 
we ignore the ac components of $\xi(t)$ and $\eta(t)$ and extract only their dc components.
This approximation is justified in $\lambda_\mu\to0$ because the time scale of change of $M(t)$ is much longer than those of  $\xi(t)$ and $\eta(t)$ and their ac components vanish by integrating Eq.~\eqref{sm:partialM2} for long time ($t_0 \gg \tau$):
\begin{align}
	M(t+t_0) 
	&= M(t) + \int_{t}^{t+t_0} ds \left[ \xi(s) - \eta(s)M(s) \right] \\
	&\sim M(t) + \int_{t}^{t+t_0} ds \left[ \xi_{0}(s) - \eta_{0}(s) M(s) \right],
\end{align}
where $\xi_{0}(t) =\sum_\mu \lambda_\mu \left[ c_\mu^+(0,t) - c_\mu^-(0,t) \right]$ and $\eta_{0}(t) = \sum_\mu \lambda_\mu \left[ c_\mu^+(0,t) + c_\mu^-(0,t) \right]$ are the dc components of $\xi(t)$ and $\eta(t)$.
Therefore, the coarse-grained solution of Eq.~\eqref{sm:partialM2} is equivalent to that of
\begin{align}
	\partial_t M(t) = \alpha(M(t)) - \beta(M(t))M(t). \label{sm:partialM3}
\end{align}
Here we have newly defined $\alpha(M(t)) = \xi_0(t)$ and  $\beta(M(t)) = \eta_0(t)$ to explicitly denote the $M(t)$-dependence:
\begin{align}
	&\alpha(M(t)) = \xi_0(t) = 2\pi \lambda_x \left[ \tgamma(\epsilon_0^+(t)) - \tgamma(\epsilon_0^-(t)) \right] + \frac{8\lambda_y}{\pi} \sum_\ell \frac{\tgamma(\epsilon_{2\ell+1}^+(t)) - \tgamma(\epsilon_{2\ell+1}^-(t))}{(2\ell+1)^2}, \\
	&\beta(M(t)) = \eta_0(t) = 2\pi \lambda_x \left[ \tgamma(\epsilon_0^+(t)) + \tgamma(\epsilon_0^-(t)) \right] + \frac{8\lambda_y}{\pi} \sum_\ell \frac{\tgamma(\epsilon_{2\ell+1}^+(t)) + \tgamma(\epsilon_{2\ell+1}^-(t))}{(2\ell+1)^2}.
\end{align}
This is the derivation of Eq.~\eqref{eq:partialM} in the main text.

\section{Floquet dynamical symmetry} \label{secap:FDS}

In this Appendix, we discuss the symmetry aspect of our DTC.
To this end, we focus on the Floquet dynamical symmetry (FDS) like Eq.~\eqref{eq:FDS0}, which plays an important role in producing the DTC~\cite{Chinzei2020}.
For instance, in isolated systems, the many-body localization protects the FDS~\eqref{eq:FDS0} and gives rise to the stable DTC~\cite{VonKeyserlingk2016}.
This FDS is generalized to dissipative systems.
Here, let us consider the following Floquet-Lindblad equation:
\begin{align}
\frac{d\rho}{dt} = \mL_t(\rho) = -i[H(t),\rho] + \sum_k \left( L_k \rho L_k^\dag - \frac{1}{2}\{L_k^\dag L_k, \rho\} \right),
\end{align}
where $\rho$ is the density matrix of the system, $H(t)=H(t+\tau)$ is the time-periodic Hamiltonian, and $L_k$'s are  quantum jump operators.
Then, the FDS is defined as~\cite{Chinzei2020},
\begin{align}
\begin{aligned}
&\hspace{1cm}U_F A U_F^\dag = e^{-i\lambda \tau}A, \\
&[L_k,A(t)] = [L_k^\dag,A(t)]=0, \quad  \forall k,t  
\end{aligned} \label{ap:FDS1}
\end{align}
where $U_F=\mathcal{T}\exp[-i\int_0^\tau{ds}H(s)]$ is the unitary one-cycle time evolution operator, $A$ and $\lambda$ are an operator and a real number characterizing the FDS, $A(t)=U(t)AU^\dag(t)$ ($ U(t)=\mathcal{T}e^{-i\int_0^t{ds}H(s)})$).
For later use, we generalize this FDS as follows:
\begin{align}
\begin{aligned}
\mU_F \mA_L = e^{-i\lambda\tau}\mA_L \mU_F, \\
\mU_F \mA_R = e^{+i\lambda\tau}\mA_R \mU_F,
\end{aligned} \label{ap:FDS3}
\end{align}
where $\mU_F=\mT \exp[\int_0^\tau{ds}\mL_s]$ is the non-unitary one-cycle time evolution superoperator, and we have defined superoperators $\mA_L(\rho)=A\rho$ and $\mA_R(\rho)=\rho A^\dag$.
Note that Eq.~\eqref{ap:FDS1} leads to Eq.~\eqref{ap:FDS3} (see supplemental material~1 in Ref.~\cite{Chinzei2020}).
This FDS~\eqref{ap:FDS3} protects the quantum coherence from the dissipation and leads to time crystalline dynamics with time scales $\tau$ and $2\pi/\lambda$.
Here, the following question naturally come to mind: Can we understand the DTC in solids in terms of the FDS?
In this Appendix, we examine this question based on the mean-field result.

Here, instead of the Lindblad equation, we consider the BR equation, $\partial_t \rho = \mR_t(\rho)$, and its non-unitary one-cycle time evolution superoperator, $\mV_F = \mT \exp[\int_0^\tau{ds}\mR_s]$.
Our mean-field theory suggests that the $\mV_F$ has an eigenvalue at $z=-1$ for the DTC phase in the thermodynamic limit, $\mV_F(\rho_{tc})=-\rho_{tc}$, which leads to the period-doubling dynamics (see also Ref.~\cite{Gong2018}). 
Also, there definitely exists an eigenvalue at $z=1$, $\mV_F(\rho_{ss})=\rho_{ss}$, due to the trace preservation.

From this spectrum, we can construct a generalized Floquet dynamical symmetry in a brute-force way.
To this end, let $\rho_{ss}^{R,L}$ and $\rho_{tc}^{R,L}$ denote left- and right-eigenstates of $\mV_F$ with eigenvalues $1$ and $-1$, respectively.
We also define a superoperator, for example, $\mA(\rho) = \rho_{tc}^R \Tr [(\rho_{ss}^L)^\dag \rho] = \rho_{tc}^R \Tr [\rho]$ (we have used $\rho_{ss}^L=1$).
Then, one obtains
\begin{align}
	\mV_F \mA = - \mA \, \mV_F, \label{ap:FDS2}
\end{align}
which can easily be shown by acting the both sides on all the right-eigenstates of $\mV_F$, $\rho_i^R$, that serve as a complete basis and using $\Tr [(\rho_i^L)^\dag \rho_j^R] = \delta_{ij}$.
This relation~\eqref{ap:FDS2} corresponds to Eq.~\eqref{ap:FDS3} with $\mU_F \to \mV_F$ and $\lambda=\pi/\tau$ although the form of $\mA$ is no longer simple as $\mA_L(\rho)=A\rho$ and $\mA_R(\rho)=\rho A^\dag$.
In summary, our DTC can also be understood from the viewpoint of the FDS, but the symmetry operator $\mA$ is so complicated that we cannot write it down in a simple form. 
Furthermore, this argument relies on the spectrum of $\mU_F$ or $\mV_F$ alone and thus would apply to any dissipative DTC.

\section{Analysis of high-frequency regime without mean-field theory} \label{secap:tau0}

Here we investigate the limit of $\tau\to0$ (i.e., $\Omega\to\infty$) without the mean-field approximation.
In this limit, the contribution of $\lambda_y$ vanishes, and only that of $\lambda_x$ remains in Eq.~\eqref{sm:partialM3}.
This can be understood by considering the original total Hamiltonian involving the system and bath:
\begin{align}
	\Htot(t) = H(t) + H_B + H_{SB},
\end{align}
where $H(t)= H_0+H_1(t)=- J \sum_{\langle i,j\rangle} \sigma^z_i \sigma^z_j + (\pi/2) \sum_{j,n} \delta(t-n\tau)\sigma^x_j$ is for the system of interest, $H_B$ is for the bath, and $H_{SB}= \sum_{j,\mu} \sqrt{\lambda_\mu} \sigma^\mu_j \otimes B^\mu_j$ is for the system-bath coupling.
Here, let us again move on to the interaction picture for $H_1(t) = (\pi/2) \sum_{n} \delta(t-n\tau)\sigma^x$:
\begin{align}
	\rho_I(t) = V^\dag(t) \rho(t) V(t),
\end{align}
with $V(t)=\mT \exp[-i\int_0^t ds H_1(s)]$.
In this picture, the density matrix $\rho_I(t)$ obeys the following von Neumann equation:
\begin{align}
	\partial_t \rho_I(t) = -i [\Htot^I(t), \rho_I(t)],
\end{align}
where $\Htot^I(t)$ is given by
\begin{align}
	\Htot^I(t) &= H_0 + H_B + H_{SB}^I(t), \\
	H_{SB}^I(t) &= \sum_{j} \sqrt{\lambda_x} \sigma^x_j \otimes B^x_j + v(t)\sum_{j,\mu=y,z} \sqrt{\lambda_\mu} \sigma^\mu_j \otimes B^\mu_j.
\end{align}
Here $v(t)$ is defined in Eq.~\eqref{sm:defv}.
Importantly, except for the terms with $\lambda_y$ and $\lambda_z$, the Hamiltonian is time-independent.
In the high-frequency limit ($\tau\to0$), since $v(t)$ oscillates rapidly, the time evolution under $\Htot^I(t)$ is identical to that under its time-average $\overline{\Htot^I}=\int_0^{2\tau} (dt/2\tau) \Htot^I(t)$~\cite{Eckardt2015}.
In other words, we can ignore the terms of $\lambda_y$ and $\lambda_z$.
Therefore, in the interaction picture, the total Hamiltonian seems time-independent, and the system relaxes to an equilibrium state due to the dissipation.
Going back to the Schr\"{o}dinger picture, we obtain the DTC state.
This is the microscopic reason why the DTC transition temperature approaches the equilibrium one in $\tau\to0$.

For small but finite $\tau$, although one cannot entirely ignore the oscillating terms that heat up the system, the dissipation could stabilize the DTC order.
In the high-frequency regime (i.e., small $\tau$), the system exhibits the Floquet prethermalization and the exponentially slow heating in $1/\tau$ according to the Floquet theory~\cite{Bukov2015}.
When the dissipation, or the system-bath coupling, is stronger than the slow heating rate,
the dissipation cools the system faster than the heating, stabilizing a prethermal DTC.
Completely elucidating whether such a prethermal DTC can exist is an open issue.

\end{widetext}


%

\end{document}